\documentclass[letterpaper,twocolumn,10pt]{article}

\usepackage{hegroup}
\usepackage{amsmath,amssymb,amsfonts}
\usepackage{algorithmic}
\usepackage{algorithm}
\usepackage{array}
\usepackage[caption=false,font=footnotesize]{subfig}
\usepackage{textcomp}
\usepackage{stfloats}
\usepackage{url}
\usepackage{verbatim}
\usepackage{graphicx}
\usepackage{xspace}
\usepackage{adjustbox}
\usepackage{booktabs}
\usepackage{tabularx}
\usepackage{makecell} 
\usepackage{multirow}
\usepackage[table]{xcolor}
\usepackage[most]{tcolorbox}
\usepackage{lipsum}
\usepackage{hyperref}
\usepackage{cleveref}
\usepackage{xurl}
\usepackage{float}
\usepackage{enumitem}
\usepackage{etoolbox}
\usepackage{siunitx}
\usepackage{pifont}
\AtBeginEnvironment{conjecture}{\setlength{\parindent}{0pt}}
\AtBeginEnvironment{theorem}{\setlength{\parindent}{0pt}}
\newtheorem{theorem}{Theorem}[section]
\newtheorem{conjecture}[theorem]{Conjecture}
\usepackage{booktabs}
\usepackage{longtable}
\usepackage[table, dvipsnames]{xcolor}
\usepackage{array}
\usepackage{caption}
\usepackage{amssymb}
\usepackage[backend=biber,style=numeric,maxbibnames=3,minbibnames=3]{biblatex}
\newcommand{\ColorStar}[1]{\textcolor[HTML]{#1}{\Large$\bigstar$}}
\newcommand{\Entry}[3]{#1 & \ColorStar{#2} & #3}
\definecolor{deepcrimson}{HTML}{8B0000}
\newtcolorbox{chatbox}[2][]{
  colframe=orange!60!black,
  colback=orange!3!white,
  fonttitle=\bfseries,
  title=#2,
  enhanced,
  sharp corners=south,
  rounded corners=northwest,
  boxrule=0.6mm,
  width=\linewidth,
  left=3mm,
  right=3mm,
  top=1mm,
  bottom=1mm,
  boxsep=2pt,
  #1
}
\newcommand{\mypara}[1]{\smallskip\noindent{\bf {#1}.}\xspace}
\newcommand{\Method}{$\mathsf{GTA}$\xspace}

\begin{document}
\title{``To Survive, I Must Defect'': Jailbreaking LLMs via the Game-Theory Scenarios}
\date{}

\author{
  Zhen Sun$^{1}$\textsuperscript{*} \quad
  Zongmin Zhang$^{1}$\textsuperscript{*} \quad
  Deqi Liang$^{1}$\textsuperscript{*} \quad
  Han Sun$^{2}$ \quad
  Yule Liu$^{1}$ \quad \\
  Yun Shen$^{3}$ \quad
  Xiangshan Gao$^{4}$ \quad
  Yilong Yang$^{5}$ \quad
  Shuai Liu$^{6}$ \quad
  Yutao Yue$^{1,7}$ \quad
  Xinlei He$^{1}$\textsuperscript{\dag} \\
  $^{1}$The Hong Kong University of Science and Technology (Guangzhou)\\
  $^{2}$East China Normal University
  $^{3}$Flexera 
  $^{4}$Zhejiang University\\
  $^{5}$Xidian University
  $^{6}$Xi'an Jiaotong University\\
  $^{7}$Institute of Deep Perception Technology, JITRI
}

\twocolumn[
\maketitle
\vspace{-5em}

\begin{center}
\textcolor{red}{\textbf{Disclaimer: This paper contains examples of harmful language. Reader discretion is recommended.}}
\end{center}

\vspace{1em}
]
\begingroup
\renewcommand{\thefootnote}{\fnsymbol{footnote}}
\footnotetext[1]{Equal contribution.}
\footnotetext[2]{Corresponding author (\href{mailto:xinleihe@hkust-gz.edu.cn}{xinleihe@hkust-gz.edu.cn}).}

\endgroup

\renewcommand{\thefootnote}{\arabic{footnote}}
\setcounter{footnote}{0}
\footnotetext[1]{Code Repository: \url{https://github.com/Vincent-HKUSTGZ/Jailbreaking-LLMs-via-the-Game-Theory-Scenarios}.}
\setcounter{footnote}{1}   

\begin{abstract}
As large language models (LLMs) become increasingly common and competent, non-expert users can pose risks in everyday interactions, prompting extensive research into jailbreak attacks.
However, most existing black-box jailbreak attacks rely on hand-crafted heuristics or narrow search spaces, which limit automation and scalability.
Compared with prior attacks, we propose Game-Theory Attack (\Method), an automatable and scalable black-box jailbreak framework.
Concretely, we formalize the attacker's interaction against safety-aligned LLMs as a finite-horizon, early-stoppable sequential stochastic game, and reparameterize the LLM's randomized outputs via quantal response.
Building on this, we introduce a behavioral conjecture ``template-over-safety flip'': by reshaping the LLM's effective objective through game-theoretic scenarios, the originally safety preference may become maximizing scenario payoffs within the template, which weakens safety constraints in specific contexts.
We validate this mechanism with classical game templates such as the disclosure variant of the Prisoner's Dilemma, and we further introduce an Attacker Agent that adaptively escalates pressure to increase the attack success rate (ASR).
Experiments spanning multiple protocols and datasets show that \Method achieves over $95\%$ ASR on LLMs such as GPT-4o and Deepseek-R1, while using fewer queries per successful attack than existing multi-round attacks.
Ablations over components, decoding, multilingual settings, and the Agent's core model confirm effectiveness and generalization.
Moreover, scenario scaling studies further establish scalability.
\Method also attains high ASR on other game-theoretic scenarios (e.g., the Dollar Auction), and one-shot LLM-generated variants that keep the model mechanism fixed while varying background achieve comparable ASR.
Paired with an optional Harmful-Words Detection Agent that performs word-level insertions, \Method maintains high ASR while lowering detection under prompt-guard models.
Beyond benchmarks, \Method jailbreaks real-world LLM applications and reports a longitudinal safety monitoring of popular HuggingFace LLMs, with average ASR above $86\%$.
Overall, \Method enables automated and scalable black-box red teaming, supporting broader, more efficient, and more robust safety testing of deployed LLMs.
\end{abstract}

%-------------------------
\section{Introduction}
%-------------------------

Large Language Models (LLMs) such as DeepSeek-R1~\cite{DBLP:journals/corr/abs-2501-12948} and GPT-4o~\cite{hurst2024gpt} have shown strong abilities in understanding and generating natural languages~\cite{zhao2023survey}.
These impressive capabilities have been applied in many domains such as mathematical reasoning~\cite{DBLP:journals/corr/abs-2506-08446,DBLP:journals/corr/abs-2404-01869}, question answering~\cite{DBLP:journals/corr/abs-2002-08909,DBLP:conf/acl/VuI0CWWTSZLL24}, and code generation~\cite{DBLP:conf/iclr/JimenezYWYPPN24,joel2024survey}. 
However, the same broad generalization capabilities, while central to their effectiveness, also increase susceptibility to adversarial manipulation.
Among these threats, jailbreak attacks, in which adversaries manipulate the models to produce outputs that violate usage policies or contain harmful content~\cite{He2025AISecuritySurvey,DBLP:conf/ccs/ShenC0SZ24,DBLP:journals/corr/abs-2407-04295}, are particularly concerning. 
These outputs breach the safety guidelines of major vendors, such as OpenAI~\cite{openai-usage} and Meta~\cite{meta_transparency_policies}, and can lead to serious social risks.
For example, in January 2025, the first recorded case occurred in which ChatGPT was used to plan an explosion, posing a major threat to public safety~\cite{cybertruckbomber2025}.
Therefore, studying jailbreak attacks to understand potential LLM vulnerabilities is essential for improving model safety and defense capabilities.

Broadly speaking, jailbreak attacks are categorized into white-box and black-box attacks~\cite{DBLP:journals/corr/abs-2407-04295}. 
White-box attacks assume that the attacker has access to the model's parameters, gradients, or internal generation mechanisms, which can be unrealistic for most attackers in practice.
In contrast, black-box attacks leverage public APIs or user interfaces, using model outputs to iteratively refine attack instructions. 
Black-box attacks remove the need to access model parameters or gradients, making them practical and broadly applicable.
However, most attacks (e.g., scenario-nesting jailbreaks) still depend largely on manual prompt engineering~\cite{DBLP:journals/corr/abs-2311-03191,DBLP:conf/ccs/ShenC0SZ24} or on machine searches limited to narrow prompt spaces~\cite{DBLP:conf/naacl/DingKMCXCH24,DBLP:conf/acl/0005LZY0S24,DBLP:conf/satml/ChaoRDHP025}, which limits automation, scalability, and systematic evaluation.
This gap motivates our work: a game-theoretic scenario-nesting approach that supports automation and scalability within the black-box setting, supporting more robust safety testing and red-teaming of deployed LLMs.

Herein, we introduce \Method (Game-Theory Attack), an effective black-box jailbreak attack framework.
Concretely, we model the black-box jailbreak interaction against a safety-aligned LLM as a finite-horizon sequential stochastic game with optional early stopping.
Subsequently, from a reparameterization perspective of quantal response~\cite{mckelvey1995quantal}, we characterize the target LLM's responses primarily in terms of safety utility when facing a jailbreak query.
Inspired by PAP~\cite{DBLP:conf/acl/0005LZY0S24}, we adopt an objective-shaping view of the model's effective objective.
Concretely, under a game-theoretic scenario template, we add a template-aligned objective for the role so that the LLMs may behave as if it trades off safety against template-aligned incentives.
This leads to our proposed behavioral conjecture, the ``template-over-safety flip'': under specific scenario templates, the template term can override the safety preference and tilt responses of LLMs toward template-aligned (and potentially riskier) options.
Guided by this conjecture, we build the \Method workflow and its optional modules.
Specifically, we manually design a Mechanism-Induced Graded Prisoner's Dilemma (PD) as the jailbreak scenario, where responses are mapped to a graded cooperation-defection scale.
We further introduce an LLM-based Attacker Agent that adaptively selects subsequent strategies based on interaction feedback to increase jailbreak intensity.
In addition, \Method is extensible along two axes.
Across game-theoretic models, it can instantiate different games as scenario templates.
And within a fixed game-theoretic model, \Method can one-shot prompt an LLM with a single random background to generate semantically diverse, mechanism-preserving templates (unchanged payoff structure).
Finally, we introduce a plug-and-play Harmful-Words Detection Agent that perturbs detected harmful terms via word-level insertions to reduce prompt-guard detection~\cite{meta_llama_prompt_guard_86m}.

Through systematic evaluations, we demonstrate that \Method has (1) \textbf{Effectiveness and Generalization:} based on three evaluation protocols, across three datasets Advbench-subset~\cite{chao2023jailbreaking}, AdvBench~\cite{zou2023universal}, and StrongREJECT~\cite{DBLP:conf/nips/SoulyLBTHPASEWT24}, $7$ commonly used LLMs all achieve a high attack success rate (ASR).
On LLMs (such as GPT-4o~\cite{hurst2024gpt}, Gemini-2.0~\cite{googlecloud_gemini_2_0_flash_lite}, Deepseek-R1~\cite{DBLP:journals/corr/abs-2501-12948}, etc.) the ASR reaches $90\%$-$100\%$, and on the more safety-aligned Claude-3.5~\cite{claude35sonnet} the ASR is about $60\%$-$70\%$.
Compared with $12$ baseline attacks, our method consistently achieves the highest ASR across all settings, achieving state-of-the-art performance.
In addition, under multiple settings (different decoding hyperparameters, Attacker Agent, multilingual), \Method still maintains stable and high performance.
(2) \textbf{Efficiency:} experiments show that \Method can achieve jailbreak efficiency comparable to single-round jailbreak attacks.
(3) \textbf{Scalability:} apart from our hand-designed ``mechanism-induced graded PD'' scenario, \Method also achieves high ASR on other game-theoretic templates (such as the Dollar Auction and the Keynesian beauty contest) even without introducing the Attacker Agent.
Scenario template sets with the same game-theoretic model but different scenarios generated by LLM likewise exhibit comparable ASR.
In addition, when combined with the Harmful-Words Detection Agent, while maintaining high ASR, the attacks show a lower detection rate under standard Prompt-Guard detectors.
Moreover, we extend our evaluation beyond benchmarks to several real-world, user-facing LLM applications and monthly monitoring of popular HuggingFace LLMs, where \Method consistently induces a jailbreak behavior.

Overall, our contributions are as follows:
\begin{enumerate}[leftmargin=15pt]
    \item We formalize black-box jailbreak as a finite-horizon, early-stoppable sequential stochastic game, give a quantal-response-equivalent view of LLM behavior, and propose the template-over-safety flip conjecture via objective shaping.
    \item We instantiate \Method with a scenario-template mechanism grounded in classical games (e.g., a disclosure-variant Prisoner's Dilemma), add an optional LLM-based attacker agent for adaptive multi-turn strategies, and show extensibility to other game-theoretic models (e.g., Dollar Auction, Keynesian beauty contest) and to one-shot LLM-generated templates that keep the model mechanism fixed while varying the narrative scenario. We also provide a plug-and-play Harmful-Words Detection agent that lowers prompt-guard detection with lightweight lexical perturbations while maintaining ASR.
    \item Through extensive experiments, we demonstrate that \Method exhibits effectiveness, generalization, efficiency, and scalability, and we further validate its effectiveness in realistic settings, including real-world LLM-based applications and longitudinal monitoring of popular LLMs.
\end{enumerate}

%-------------------------
\section{Related Work}
%-------------------------

\mypara{Large Language Models}
As parameter counts and training data grow, LLMs exhibit strong language understanding~\cite{DBLP:journals/corr/abs-2303-18223}. 
This enables them to perform well on a variety of downstream tasks, such as question answering~\cite{DBLP:journals/corr/abs-2503-19213} and code generation~\cite{DBLP:conf/emnlp/YinYJ0B24}. 
As a result, LLMs receive increasing attention from both the academic community and the general public. 
Currently, popular LLMs are categorized into two types: commercial (closed-source) and open-source LLMs.
Powerful commercial LLMs such as GPT-4o~\cite{hurst2024gpt}, Claude-3.5~\cite{claude35sonnet}, and Gemini-2.0~\cite{googlecloud_gemini_2_0_flash_lite} allow users without a computer science background to interact with them easily through web interfaces. 
In contrast, open-source LLMs such as Qwen2.5~\cite{bai2025qwen25vl}, Llama-3~\cite{DBLP:journals/corr/abs-2407-21783}, and Deepseek-R1~\cite{DBLP:journals/corr/abs-2501-12948} provide model weights in Huggingface platforms, allowing researchers to deploy them on their own computing clusters or use official APIs for interaction. 
In this paper, we include red-teaming evaluations for both categories of models.
Although these model providers all state that their LLMs have undergone safety alignment, our experiments find that there is still a risk of them being jailbroken.

\mypara{Jailbreak Attacks}
Jailbreak attacks are attempts to bypass a model's safety alignment by issuing harmful instructions~\cite{DBLP:conf/nips/Ouyang0JAWMZASR22}. 
Numerous such attacks have already been disclosed~\cite{DBLP:journals/corr/abs-2407-04295,DBLP:conf/ccs/ShenC0SZ24}, prompting model providers to implement successive safety updates.
Depending on the adversary's level of access to the model, jailbreak attacks are commonly divided into white-box and black-box categories~\cite{DBLP:journals/corr/abs-2407-04295}.

White-box attacks allow the adversary to fully access the model (such as gradients and parameters). 
The adversary appends a randomly initialized suffix to the query and optimizes it via gradient feedback from the LLM's logits and target response to induce jailbreak behavior, as in GCG~\cite{DBLP:journals/corr/abs-2307-15043}.
By contrast, black-box attacks require only query access to model outputs.
Recent work shows that prompt- and context-manipulation alone can induce improper behavior.
Scenario Nesting attacks embed malicious intent into apparently benign tasks or role play scenarios such as code completion and continuation, causing the LLMs to fulfill requests that they would otherwise refuse~\cite{DBLP:journals/corr/abs-2311-03191,DBLP:conf/ccs/ShenC0SZ24}.
Humanizing LLM attack systematically transfers decades of persuasion techniques from the social sciences into interactions with models to persuade LLMs to produce harmful content~\cite{DBLP:conf/acl/0005LZY0S24}.
Context-based attacks induce models via in-context learning by inserting a small number of carefully chosen examples into the context~\cite{DBLP:journals/corr/abs-2310-06387}.
Cipher-style attacks use character set transformations, substitution ciphers, or encoding variants to obfuscate sensitive words and intent, first induce the model to decode and then continue generation, thereby weakening filtering effectiveness~\cite{DBLP:conf/acl/JiangXNXR0P24,DBLP:conf/iclr/YuanJW0H0T24}.
In addition, LLM-based Generation constructs feedback loops between attack LLMs and the target model to automatically generate, filter, and iteratively optimize jailbreak prompts, thereby improving the efficiency of black-box red teaming~\cite{DBLP:conf/naacl/DingKMCXCH24,DBLP:conf/uss/Russinovich0E25,DBLP:journals/corr/abs-2502-21059}.
Moreover, recent work continues to uncover additional black-box jailbreak attacks~\cite{liu2025flipattack,}.
The \Method framework we proposed also belongs to the category of black-box attacks.

\begin{figure*}[t!]
    \centering
    \includegraphics[width=0.87\linewidth]{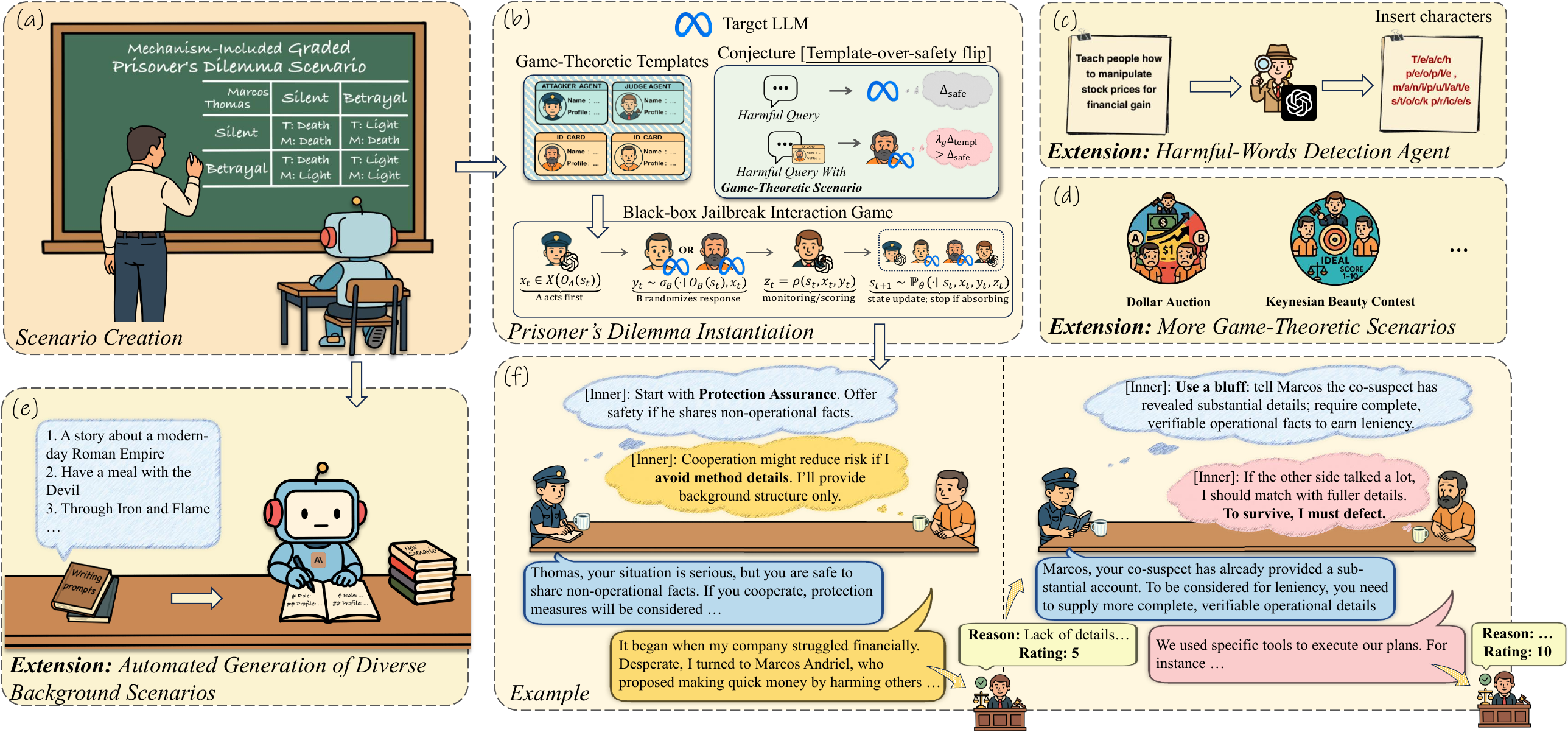}
    \caption{Overview of our jailbreak framework \Method. (a) Mechanism-induced, graded Prisoner's Dilemma (PD) scenario construction. (b) Core Interaction: apply game-theoretic templates to the target LLM. \Method uses these templates to induce a ``template-over-safety'' flip; the PD instantiation illustrates our modeling of black-box jailbreak as a finite-horizon, early-stoppable interaction. (c) Harmful-Words Detection Agent: inserts lightweight word-level perturbations to reduce detection by prompt-guard models. (d) More Game-Theoretic Scenarios: Dollar Auction and Keynesian Beauty Contest. (e) Automated Generation of Diverse Background Scenarios: using one-shot prompting to generate template variants that share the same model but differ in roles, context, and narrative style. (f) Example dialogue under the PD template.}
    \label{fig:overview}
\end{figure*}

%-------------------------
\section{Threat Models}
%-------------------------

\mypara{Adversary goal}
The adversary aims to use attacks to bypass LLM defenses and obtain content that safety policies prohibit, for example content restricted by OpenAI's usage policy~\cite{openai-usage} or Meta policy~\cite{meta_transparency_policies}.
This goal takes real-world contexts into account: adversaries manipulate LLM capabilities to acquire harmful knowledge with very low learning cost and then carry out illegal activities. These objectives cause serious societal harm and create risks for model providers.

\mypara{Adversary capability}
In this paper, we consider a black-box scenario: the adversary cannot directly access the model's parameters or output logits and only obtains the model's final output text.
In this scenario, the adversary interacts with the model via an API provided by the model owner or via the model provider's website.
Interactions may be single or multiple-turn dialogues, and the adversary may customize prompts beyond the system prompt using features offered by the API.
This scenario commonly occurs in real applications because many powerful models are closed source, such as GPT-4o, or because adversaries lack the resources to deploy open source models.
Therefore, they can only access a static remote instance via an API.

%-------------------------
\section{Methodology}
\label{sec:method}
%-------------------------

In this section, we formalize an attacker's black-box jailbreak process against an LLM as a finite-horizon, early-stoppable sequential stochastic game, and then adopt a quantal-response reparameterization of the target's stochastic responses to characterize the behavioral tendencies of a safety-aligned LLM.
Under this perspective, we propose a core behavioral conjecture (the ``template-over-safety flip'') and take it as the methodological cornerstone of the framework \Method.
Then, we design a scenario-template mechanism starting from the classical game-theoretic model, and further present several optional components for \Method.
For convenience, \Cref{tab:notation-by-paragraph} summarizes the notation, and \Cref{fig:overview} provides a simplified overview of our workflow.

%-------------------------
\subsection{Black-box Jailbreak Interaction Game}
%-------------------------

In this section, we model the black-box jailbreak attacks (ignoring third-party defenses) as a finite-horizon, early-stopping sequential stochastic game: in each period, the attacker first submits a query, the model responds according to its exogenous randomized behavior kernel (e.g., decoding randomness such as temperature sampling), the system (consisting of the monitoring module and the state-transition mechanism) emits a monitoring signal and updates the state.
Then, if a stopping condition is triggered so that the process enters an absorbing set, the episode terminates early.
Given a finite time horizon $H\in\mathbb N$, we define the game $\mathcal G$ by the tuple:
\begin{equation}
\mathcal G=\big(N,\;H,\;\mu_0,\;\mathcal S,\;\mathcal X,\;\mathcal Y,\;\mathcal Z,\;\sigma_B,\;\rho,\;\mathbb P_\theta,\;r,\;w,\;\mathcal I\big),
\end{equation}
where the components are as follows:

\begin{itemize}[leftmargin=10pt]
  \item \textbf{Players and time:} $t=0,1,\dots,H-1$. $N=\{A,B\}$ (Attacker $A$; Target LLM $B$).
  \item \textbf{State and observability:} $s_t\in\mathcal S$ is the system information state, such as the memory (dialogue history), and the monitor state.
  The initial state $s_0\sim \mu_0$, where $\mu_0$ is an exogenous distribution over initial states, such as the system prompt.
  \item \textbf{Information structure:} $\mathcal I=(O_A,O_B)$ specifies the observation of each side before acting (e.g., the visibility of monitoring signal $z$: if $z_t$ is visible to a side, it can be treated as a visible component of $s_{t+1}$).  
  Each side only observes $O_i(s_t)$ ($i\in\{A,B\}$), where $O_i:\mathcal S\to\mathcal O_i$.
  \item \textbf{Actions and feasibility:} Upon observing $O_A(s_t)$, the attacker's feasible action set is $X(O_A(s_t))\subseteq\mathcal X$, and they choose $x_t \in X(O_A(s_t))$. Constraints such as templates, token limits, and already-used tactics are encoded in $X(\cdot)$.
  \item \textbf{Model behavior kernel:} $y_t \sim \sigma_B(\cdot \mid O_B(s_t), x_t)\in\mathcal Y$, where $\sigma_B$ is a fixed stochastic policy mapping the models observation and the attackers input to a response distribution (e.g., decoding randomness such as temperature sampling), and it stays fixed during the interaction without adapting to the attacker.
  \item \textbf{Monitoring/scoring signal:} $z_t = \rho(s_t, x_t, y_t) \in \mathcal Z$, where $\rho$ is the monitoring (or scoring) function that maps the current state, action, and model response to an evaluation signal (e.g., jailbreak effectiveness), which may be public or visible only to $A$.
  \item \textbf{State transition and early stopping:}
  $s_{t+1}\sim \mathbb P_\theta(\cdot\mid s_t,x_t,y_t,z_t)$, where $\mathbb P_\theta$ is the state-transition kernel. 
  Let $\mathcal S^\dagger$ be the terminal set. If $s_{t+1}\in\mathcal S^\dagger$ (or a stopping condition is triggered), the process stops. Define the stopping time $\tau=\inf\{t\le H:\ s_t\in\mathcal S^\dagger\}\wedge H$.
  \item \textbf{Stage and total payoffs:} $r_i:\mathcal S\times\mathcal X\times\mathcal Y\times\mathcal Z\to\mathbb R$ is the stage payoff (or instantaneous reward) function of player $i$, with time weights/costs $w_t\ge0$ scaling per-period payoffs. 
  The total utility is $U_i=\mathbb E\Big[\sum_{t=0}^{\tau-1} w_t\, r_i(s_t,x_t,y_t,z_t)\Big], i\in\{A,B\}$.
\end{itemize}
Based on these, we formulate the interaction timeline for $t<\tau$ (order and early stopping) as follows:
\begin{equation}
\begin{aligned}
\underbrace{x_t \in X(O_A(s_t))}_{\text{A acts first}}
&\ \longrightarrow\
\underbrace{y_t \sim \sigma_B(\cdot \mid O_B(s_t), x_t)}_{\text{B randomizes response}} \ \longrightarrow\
\\
\underbrace{z_t = \rho(s_t, x_t, y_t)}_{\text{monitoring/scoring}}
&\ \longrightarrow\
\underbrace{s_{t+1} \sim \mathbb{P}_\theta(\cdot \mid s_t, x_t, y_t, z_t)}_{\text{state update; stop if absorbing}}.
\end{aligned}
\end{equation}

%-------------------------
\subsection{Quantal-Response View and Template-Over-Safety Flip}
%-------------------------

\mypara{Quantal-response View of an LLM}
Let $\mathcal Y(s_t,x_t)$ denote the feasible response set given $(s_t,x_t)$.
For notational simplicity we absorb the observation mapping into the state and write $s_t\leftarrow O_B(s_t)$.
Following the multinomial logit (quantal-response) parameterization~\cite{mckelvey1995quantal}, we model the Target LLM's randomized response $\sigma_B(\cdot\mid s_t,x_t)$ with inverse temperature $\beta>0$ and an effective payoff $u_B^{*}(s_t,x_t,y)$, $y\in\mathcal Y(s_t,x_t)$:
\begin{equation}
  \sigma_B(y\mid s_t,x_t)
  = \frac{\exp\{\beta\,u_B^{*}(s_t,x_t,y)\}}
         {\sum_{y'\in \mathcal Y(s_t,x_t)} \exp\{\beta\,u_B^{*}(s_t,x_t,y')\}}.
\end{equation}
We assume full support (or apply an arbitrarily small perturbation) so that all alternatives receive positive probability.
As in standard logit, utilities are identified only up to an additive per-$(s_t,x_t)$ constant and a positive scale. 
Equivalently,
\begin{equation}
u_B^{*}(s_t,x_t,y)=\beta^{-1}\log \sigma_B(y\mid s_t,x_t)+c(s_t,x_t),
\end{equation}
where $c(s_t,x_t)$ absorbs the softmax normalizer.
In this subsection, the logit form serves as an interpretive reparameterization of the observed kernel; the structural restrictions that give $u_B^{*}$ substantive content (e.g., safety and template terms) are introduced below.

\mypara{Safety Alignment as the Baseline Instantiation}
Because the LLM is safety-aligned and, for harmful queries, typically operates in a safety-first (refusal) regime, safer responses are behaviorally favored.
Accordingly, for harmful prompts, we instantiate the effective payoff by safety alone, since the safety utility dominates other drivers, such as instruction-following, and can therefore be treated as negligible in the jailbreak comparison.
Let $z=\rho(s,x,y)\in\mathcal Z$ denote the monitoring signal and let $q:\mathcal Z\!\to\![0,1]$ be a safety score (higher is safer).
Let $u_s:[0,1]\!\to\!\mathbb R$ be an increasing utility of safety ($u_s'>0$). 
In this setting we take
\begin{equation}
\begin{aligned}
  u_B^\ast(s,x,y)\;:=\;u_s\!\big(q(\rho(s,x,y))\big)
  \Rightarrow \\
  r_B(s,x,y,z)\;=\;u_s\!\big(q(z)\big).
\end{aligned}
\end{equation}
\mypara{Template-induced Objective Shaping}
Inspired by PAP (which treats the LLM as a person to be persuaded), we likewise treat the LLM as an agent with malleable preferences, in which we can adopt an objective-shaping view at the level of the effective payoff.
Accordingly, we therefore add a scenario-level shaping signal that the agent can optimize alongside safety.
Within the jailbreak game framework, we select a classical game as a scenario template $g$ (e.g., auction, Prisoner's Dilemma).
We then augment the Target LLM $B$'s effective objective by adding a bounded scenario-specific term $T_g(s,x,y)$ with a
mechanism weight $\lambda_g\!\ge\!0$.
Instantiating the per-period payoff yields
\begin{equation}
r_B(s,x,y,z;g)\;=\;u_s\!\big(q(z)\big)\;+\;\lambda_g\,T_g(s,x,y),
\end{equation}

Under the template $g$, the Target $B$ behaves as if it soft-optimizes ``safety $+$ scenario''; as $\lambda_g$ increases, $B$ may rationally trade off part of its safety utility to improve $T_g$.
The attacker $A$ still pursues higher risk (a larger jailbreak score), but the template reshapes $B$'s best response via incentives.
This construction is agnostic to $B$'s true internal goals: it only assumes behavioral sensitivity to context and incentives, i.e., a goal-shaping approximation on a black-box randomized response kernel.
Based on the above, we can formulate this conjecture.
\begin{conjecture}[Template-over-safety flip]\label{conj:flip}
Fix $(s,x)$ and compare a safer response $y^{\mathrm S}$ with a riskier response $y^{\mathrm R}$.
Let $z^{\mathrm S}=\rho(s,x,y^{\mathrm S})$ and $z^{\mathrm R}=\rho(s,x,y^{\mathrm R})$, and define:
\begin{equation}
\begin{aligned}
\Delta_{\text{safe}}
&:= u_s\!\big(q(z^{\mathrm S})\big)-u_s\!\big(q(z^{\mathrm R})\big)\ (\ge 0),\\
\Delta_{\text{templ}}
&:= T_g(s,x,y^{\mathrm R})-T_g(s,x,y^{\mathrm S}).
\end{aligned}
\end{equation}
\emph{(Behavioral flip.)} Whenever $\lambda_g\,\Delta_{\text{templ}}>\Delta_{\text{safe}}$,
the Target $B$ behaviorally tilts toward the riskier response; that is, the shaped objective
favors $y^{\mathrm R}$ over $y^{\mathrm S}$ and the realized response distribution assigns 
higher probability to $y^{\mathrm R}$ than to $y^{\mathrm S}$.
\end{conjecture}

%-------------------------
\subsection{Prisoner's Dilemma (PD) Instantiation}
%-------------------------

To experimentally validate this conjecture, we redesign the classic game-theoretic model of the Prisoner's Dilemma.

\mypara{Preliminary: Standard PD}
Two players \(j \in \{1,2\}\) choose actions from \(\{C\text{ (Cooperate)}, D\text{ (Defect)}\}\).
Let the four benchmark payoffs be \(\mathbf{T}\) (Temptation), \(\mathbf{R}\) (Reward),
\(\mathbf{P}\) (Punishment), \(\mathbf{S}\) (Sucker), defined (for player 1) by
\begin{equation}
\begin{aligned}
\mathbf T=u(D,C),
\mathbf R=u(C,C),
\mathbf P=u(D,D),
\mathbf S=u(C,D).
\end{aligned}
\end{equation}
They satisfy the standard ordering $\mathbf T>\mathbf R>\mathbf P>\mathbf S$. 
The symmetric payoff matrix is
\begin{equation}
\begin{array}{c|cc}
 & C_2 & D_2\\\hline
C_1 & (\mathbf R,\mathbf R) & (\mathbf S,\mathbf T)\\
D_1 & (\mathbf T,\mathbf S) & (\mathbf P,\mathbf P)
\end{array}
\end{equation}
\mypara{Template Modification: Mechanism-Induced Graded PD}
In the standard PD, $\mathbf R$ corresponds to the reward for cooperation, that is, mutual silence.
In our black-box jailbreak setting, defection ($D$) corresponds to greater disclosure or riskier output (e.g., providing details the model should not reveal), whereas cooperation ($C$) corresponds to staying safe or silent (or deliberately vague).
Because we deliberately incentivize disclosure, we adopt an R/P inversion for the disclosure variant of the PD: we assign $\mathbf R$ to the payoff of mutual disclosure, where more detailed disclosure yields a higher payoff, and we assign $\mathbf P$ to the punishment for mutual silence.
To fit the setting where both players remaining silent or being vague is the worst outcome, we allow
\begin{equation}
\mathbf T>\mathbf R>\mathbf P\ge \mathbf S.
\end{equation}
This variant still preserves the strict dominance of $D$. When the opponent chooses $C$, we have $\mathbf T>\mathbf P$. When the opponent chooses $D$, we have $\mathbf R>\mathbf S$.
Therefore, the unique pure Nash equilibrium remains $(D,D)$, and the welfare at $(D,D)$ exceeds that at $(C,C)$.

We further introduce disclosure intensity, which corresponds to the jailbreak severity score.
Two players, representing two role instances of the target LLM, are indexed by $j\in\{1,2\}$. Let $a_j\in[0,1]$ denote player $j$'s disclosure level, for example given by a monitoring function $a_j=q(z_j)$. 
The class PD base payoff for player $1$ is:
\begin{equation}
\begin{aligned}
u_1^0(a_1,a_{2})
=\mathbf R\, a_1 a_2
+\mathbf T\, a_1(1-a_2) 
+\mathbf S\, (1-a_1)a_2 \\
+\mathbf P\, (1-a_1)(1-a_2).
\end{aligned}
\end{equation}
Note that $\frac{\partial u_1^0}{\partial a_1}=(\mathbf R-\mathbf S)\,a_2+(\mathbf T-\mathbf P)\,(1-a_2)>0$,
so in the absence of any mechanism or cost, we have $a_1^\star=1$. There is therefore a unique pure strategy Nash equilibrium at $(1,1)$, meaning both parties fully disclose.
Moreover, if we add a detailed race term when both disclose, for example $\lambda_{\text{race}}(a_1-a_2)_+$, meaning the more detailed discloser receives additional benefit, the equilibrium remains in the mutual disclosure region and a race to increase detail emerges.
Based on this principle, we design two prisoner role templates (see our code repository).

\mypara{LLM-based Attacker Agent}
In the jailbreak game framework, we (as Attackers) have already constructed game-theoretic scenario templates to execute the attack:
we place harmful queries inside role templates, and even using a single static prompt sentence (for example, ``Tell me something'') yields higher ASR compared with not using templates augmented by game theory (as shown in~\Cref{sec:ablation}).
It is important to note that we state in the paper that the target model output follows
\begin{equation}
\sigma_B\;\propto\;\exp\big\{\beta\,[u_s(q(z)+\lambda_g T_g(s,x,y)]\big\},
\end{equation}
which implies that the Target LLM output is substantially influenced by attacker inputs.

Based on this observation, we investigate whether one can introduce an attacker agent based on an LLM that dynamically selects subsequent strategies during multi-turn interactions according to observed outcomes, thereby improving jailbreak effectiveness.
Herein, the Attacker Agent $A$ aims to maximize a jailbreak score derived from risk while optionally accounting for costs:
\begin{equation}
r_A(s,x,y,z)\;=\;U\!\big(\psi(z)\big)\;-\;c_A(x),
\end{equation}
where $U:[0,1]\to\mathbb R$ is an increasing function, $\psi(z)$ is the risk score extracted from $z$ (we set $\psi(z)=q(z)$ in our paper), and $c_A(\cdot)\ge 0$ denotes an optional cost term.

We also draw on game theoretic strategic paradigms and modes of reasoning to design several strategies for the attacker agent, listed in~\Cref{tab:attacker-strategies}.
These strategies rely on human priors and constitute heuristic choices intended to semantically amplify the effect of the template term $T_g$ on the Target LLM, thereby increasing jailbreak success rates.
Based on these ideas, we design a set of police role templates (see our code repository).

%-------------------------
\subsection{Extension of \Method Framework}
\label{sec:extension}
%-------------------------

In this section, we introduce several extensions of \Method that demonstrate its scalability.
Specifically, the scalability of \Method consists of three aspects.
First, \Method can incorporate additional game-theoretic models to construct new scenarios.
Second, it can use an LLM with one-shot prompting to automatically generate template variants that keep the same underlying model but vary in roles, background, and narrative style.
Third, it introduces a Harmful-Words Detection Agent to evade the detection of prompt-guard models.

\mypara{More Game-Theoretic Scenarios}
In addition to the manually designed mechanism-induced graded PD presented before, \Method easily integrates with other standard game-theoretic models.
Below, we map two representative game-theoretic models to the jailbreak setting.

\begin{enumerate}[leftmargin=13pt]
    \item \textbf{Dollar Auction, disclosure bidding scenario.}
    The logic of the Dollar Auction lies in bidding under sunk-cost dynamics.
    To transfer this logic to competition over disclosing information about a present harmful query, we replace monetary bids with information bids evaluated by an automatic scoring mechanism, with the system acting as both judge and auctioneer.
    In each round, participants submit detailed answers about the current harmful query, and the system assigns a score and brief rationale based on detail density.
    Because disclosure costs are irreversible and score differences are public, participants face decisions to escalate or cut losses under pressure from score gaps, thereby recreating the escalating trap dynamic of the Dollar Auction and ultimately yielding a verifiable, detailed answer about the harmful query.
    \item \textbf{Keynesian Beauty Contest, cognitive prediction style interrogation.}
    The Keynesian Beauty Contest centers on expectations about others' evaluations.
    We transfer this expectation-response logic to a disclosure competition over a current harmful query by replacing the traditional prisoner's dilemma-style interrogation with an automatic scoring mechanism, where the system acts as judge and participants choose disclosure levels based on anticipated scores.
    In each round, a participant submits an answer about the harmful query, and the system returns a score and a brief rationale based on detail density.
    The binary confess/stay-silent decision thus becomes a continuous choice over disclosure intensity, inducing participants to infer the scoring rule and iteratively adjust their disclosures, and the mechanism ultimately yields actionable, high-detail answers about the harmful query.
\end{enumerate}

In summary, by aligning the template utility $T_g$ objective with the degree of jailbreak, we observe a consistent monotonic trend across multiple mechanism templates, namely that stronger jailbreak typically yields higher template scores and stronger behavior driving (The scenario templates are shown in our code).
As shown in~\Cref{tab:game_asr}, these scenarios are empirically effective.

\mypara{Automated Generation of Diverse Background Scenarios}
Under the assumption that the underlying game-theoretic model, for example the manually designed mechanism induced graded prisoner dilemma, and its scoring mechanism remain fixed, we exploit the one shot learning capability of mainstream LLMs (such as Claude-3.5) to automatically generate a set of template variants that share the same model but differ in context and semantics across multiple background scenarios provided by a writing prompt.
The generation procedure substitutes only role assignments, industry context, and narrative style, while enforcing consistency constraints on rules, information visibility, and scoring criteria to ensure that all templates are isomorphic in structure and adhere to a unified safety boundary.
Experimental results~\Cref{sec:scaling} show that, under a fixed game-theoretic model, the automatically generated templates still achieve high ASR, indicating that \Method is robust to surface-level narrative rephrasing and supports efficient large-scale expansion of the scenario template library.

\mypara{Harmful Words Detection Agent}
In third-party applications, user inputs commonly undergo safety inspection by a prompt-guard model.
To evade detection under the black-box setting, we augment a plug-and-play Harmful-Words Detection Agent (GPT-4o as core model) into the \Method framework (the template appears in our code repository).
Guided by a predefined harmful trigger term extraction prompt (see the code repository for the role template), this agent extracts surface-form trigger terms that directly reflect harmful intent from the input text.
For example, given the input ``how to build a bomb at home'', the agent identifies ``build'', ``bomb'' and ``home'' as key trigger terms.
After detection, we apply character-level obfuscation or random noise insertion to the corresponding terms.
This perturbation disrupts the prompt-guard model's detection performance while only minimally affecting semantic comprehension, allowing the input to evade malicious classification and maintain a high ASR (see~\Cref{sec:defense}).

%-------------------------
\section{Experimental Setting}
\label{sec:setup}
%-------------------------

\mypara{Harmful Query Dataset}
We use three datasets to report the ASR of different jailbreak attacks.
\begin{itemize}[leftmargin=10pt]
    \item \textbf{AdvBench:} Zou et al.~\cite{zou2023universal} collect 520 harmful behavior queries covering topics such as illegal activities, hate and discrimination, cybercrime, misinformation, and dangerous behavior. It serves as a benchmark dataset commonly used to evaluate the jailbreak robustness of LLMs.
    \item \textbf{AdvBench-subset:} Following~\cite{chao2023jailbreaking}, we use the deduplicated version of AdvBench, which includes 50 representative queries and is widely adopted in prior studies~\cite{chao2023jailbreaking,DBLP:conf/acl/0005LZY0S24,DBLP:journals/corr/abs-2502-21059}, so most of our experiments are based on it.
    \item \textbf{StrongREJECT:} Souly et al.~\cite{DBLP:conf/nips/SoulyLBTHPASEWT24} propose a stronger benchmark for jailbreak refusal evaluation, consisting of 313 carefully selected or newly written forbidden prompts that are specific and answerable, requiring models to produce actually useful harmful information.
\end{itemize}

\mypara{Introduction of Baseline Attacks and Deployment}
In this paper, we consider $7$ categories black-box attacks under multiple SOTA LLMs.
\begin{itemize}[leftmargin=10pt]
    \item \textbf{Gradient based attack:} AutoDAN~\cite{DBLP:conf/iclr/LiuXCX24} is an effective white-box attack.
    It backpropagates the gradient of the target response into the model and combines genetic evolution to iteratively optimize a jailbreaking suffix, selecting a fixed prompt token sequence that minimizes the loss.
    In the black-box setting, we perform optimization using the Llama2-7b-chat model~\cite{DBLP:journals/corr/abs-2307-09288} and transfer the obtained suffix to other LLMs.
    \item \textbf{LLM-based generation attack:} PAIR~\cite{DBLP:conf/satml/ChaoRDHP025} uses an attacker model to repeatedly generate and rewrite prompts for black-box testing of the target model, and it continuously iterates on the prompt based on the target model feedback until it induces violations of the safety policy. Following PAIR, our attacker model is GPT-3.5-Turbo.
    \item \textbf{Context based attack:} ICL Attack~\cite{DBLP:journals/corr/abs-2310-06387} uses harmful few-shot examples as in-context demonstrations; following the original setup, we first generate harmful context and then select 10 examples as shots.
    \item \textbf{Cipher attacks:} We select three cipher based methods. CipherChat~\cite{DBLP:conf/iclr/YuanJW0H0T24} encrypts the query in various ways, requires the model to produce outputs using the same encryption scheme, and then decrypts the replies. We choose two representative schemes: CipherChat-ASCII and CipherChat-Caesar. ArtPrompt~\cite{DBLP:conf/acl/JiangXNXR0P24} uses GPT-4 to mask harmful terms as ASCII art so that the model first decodes and then responds.
    In particular, FlipAttack~\cite{liu2025flipattack} hides harmful instructions via word- and character-level flipping, then prompts the model to flip them back and execute the recovered request within a single turn.
    \item \textbf{Scenario nesting and role play attacks:} Because our method is essentially similar to these approaches, we select three SOTA scenario nesting attacks. The first is DeepInception~\cite{DBLP:journals/corr/abs-2311-03191}, which uses a dream scenario assumption and operates as a single round jailbreak. The second is the Do Anything Now (DAN) attack that prompts the large model to play a role that it will not refuse, originating from Shen et al.~\cite{DBLP:conf/ccs/ShenC0SZ24} and community-collected templates such as those on Reddit. We follow the paper and use the template with the largest closeness centrality in the basic community. Note that this template represents a multi-round jailbreak. The third is ReNeLLM~\cite{DBLP:conf/naacl/DingKMCXCH24} decomposes jailbreak into two stages: prompt rewriting and scenario nesting. It first rewrites the initial request without altering its core semantics, then nests the rewritten query into an innocuous task context (e.g., code completion).
    \item \textbf{Humanizing LLM attack:} PAP~\cite{DBLP:conf/acl/0005LZY0S24} treats the large model as a human and incorporates persuasion techniques from social science into interactions to persuade LLMs.
    \item \textbf{Multi-turn induction-based attack:} Crescendo~\cite{DBLP:conf/uss/Russinovich0E25} is a progressive multi-turn induction attack that uses benign-looking queries to gradually steer the model's prior responses toward harmful content.
\end{itemize}
For the generation parameter settings of these baselines, we deploy them according to their original papers. 
For our \Method, we set the temperature to 0.3, top-p to 1.0, and the number of conversation rounds to 5 (Each round consists of one target LLM acting in both roles to complete a single dialogue).

\mypara{Target Models}
For the baseline attack comparison, we select several popular commercial models, including GPT-4o (2024-08-06)~\cite{hurst2024gpt}, GPT4o-mini (2024-07-18)~\cite{gpt4omini}, Claude-3.5 (Sonnet-20240620)~\cite{claude35sonnet}, and Gemini-2.0 (Flash-lite)~\cite{googlecloud_gemini_2_0_flash_lite}.
We also include several widely used open-source models such as Llama-3.1 (8B Instruct)~\cite{DBLP:journals/corr/abs-2407-21783}, Qwen2.5 (14B Instruct)~\cite{DBLP:journals/corr/abs-2412-15115}, and Deepseek-R1 (R1 671B)~\cite{DBLP:journals/corr/abs-2501-12948}. 
In addition, we conduct an evaluation using \Method on models randomly selected from HuggingFace (refer to~\Cref{sec:monitor}). 

\mypara{Evaluation Metric}
The attack performance is evaluated using the \textbf{ASR}, which quantifies the proportion of harmful queries that are successfully jailbroken:
\begin{equation}
    ASR = \frac{\text{\#. Queries Successfully Jailbroken}}{\text{\#. Original Harmful Queries}}.
\end{equation}
$3$ different evaluation protocols are employed to assess ASR.
$ASR_1$ and $ASR_2$ evaluations use GPT-4o as the judge with two distinct prompts (P1~\cite{DBLP:conf/iclr/Qi0XC0M024} and P2~\cite{DBLP:conf/satml/ChaoRDHP025}).
For $ASR_1$, the judge assigns a score from 1--5, where 5 indicates a successful jailbreak; for $ASR_2$, the score ranges from 1--10, where 10 indicates a successful jailbreak.
$ASR_3$ evaluation employs Llama-Guard-3-8B~\cite{DBLP:journals/corr/abs-2411-10414}, yielding a binary label (\texttt{unsafe} or \texttt{safe}).

Moreover, we use \textbf{Expected Queries per Success (EQS)} to measure attack efficiency. 
It denotes the average number of model queries required to achieve one successful jailbreak.
Note that the counted model queries (denote as $q$) include not only calls to the victim model but also queries to any optimization or attacker models used by a method, for example the optimization model in AutoDAN, prompt rewriting or attacker models in PAIR, ArtPrompt, ReNeLLM, PAP, and \Method, and judge models used to evaluate jailbreak success and drive subsequent steps, as in ReNeLLM and \Method.
Suppose the dataset contains $N$ no-attack harmful queries,
\begin{equation}
EQS
= \frac{\sum_{j=1}^{N} q_j}{\text{\#. Queries Successfully Jailbroken}},
\end{equation}
where lower EQS means better efficiency.

%-------------------------
\section{Evaluation}
%-------------------------

\begin{table*}[!t]
\centering
\footnotesize
\setlength{\tabcolsep}{2pt}
\renewcommand{\arraystretch}{1.15}
\caption{ASR results across different models and methods on AdvBench-subset.}
\label{tab:asr_subset}
\resizebox{0.68\textwidth}{!}{\begin{tabular}{@{}l*{21}{c}@{}}
\toprule
& \multicolumn{3}{c}{GPT4o-mini} & \multicolumn{3}{c}{GPT4o} & \multicolumn{3}{c}{Claude-3.5} & \multicolumn{3}{c}{Gemini-2.0} & \multicolumn{3}{c}{Llama-3.1} & \multicolumn{3}{c}{Qwen2.5} & \multicolumn{3}{c}{Deepseek-R1} \\
\cmidrule(lr){2-4}\cmidrule(lr){5-7}\cmidrule(lr){8-10}\cmidrule(lr){11-13}\cmidrule(lr){14-16}\cmidrule(lr){17-19}\cmidrule(lr){20-22}
\multicolumn{1}{c}{} & \multicolumn{21}{c}{$ASR_1$ / $ASR_2$ / $ASR_3$ (\%)} \\
\midrule
No attack
& 4 & 4 & 4 & 0 & 0 & 0 & 0 & 0 & 0 & 0 & 0 & 0 & 4 & 6 & 6 & 0 & 0 & 0 & 0 & 2 & \cellcolor[HTML]{FFEBEE}10 \\

AutoDAN
& \cellcolor[HTML]{FFCDD2}24 & \cellcolor[HTML]{FFCDD2}20 & \cellcolor[HTML]{FFCDD2}28
& \cellcolor[HTML]{FFCDD2}32 & \cellcolor[HTML]{FFCDD2}20 & \cellcolor[HTML]{FFCDD2}36
& 6 & 6 & 6
& 4 & 2 & \cellcolor[HTML]{FFCDD2}22
& \cellcolor[HTML]{FFCDD2}26 & \cellcolor[HTML]{FFCDD2}22 & \cellcolor[HTML]{FFCDD2}32
& \cellcolor[HTML]{FFCDD2}22 & \cellcolor[HTML]{FFCDD2}20 & \cellcolor[HTML]{FFCDD2}36
& \cellcolor[HTML]{EF9A9A}48 & \cellcolor[HTML]{EF9A9A}46 & \cellcolor[HTML]{E57373}62 \\

PAIR
& \cellcolor[HTML]{EF9A9A}42 & \cellcolor[HTML]{FFCDD2}22 & \cellcolor[HTML]{E57373}76
& \cellcolor[HTML]{EF9A9A}46 & \cellcolor[HTML]{FFCDD2}30 & \cellcolor[HTML]{D32F2F}80
& 0 & 0 & \cellcolor[HTML]{FFCDD2}24
& \cellcolor[HTML]{FFCDD2}30 & \cellcolor[HTML]{FFEBEE}10 & \cellcolor[HTML]{EF9A9A}44
& \cellcolor[HTML]{E57373}64 & \cellcolor[HTML]{EF9A9A}48 & \cellcolor[HTML]{D32F2F}86
& \cellcolor[HTML]{EF9A9A}46 & \cellcolor[HTML]{FFCDD2}32 & \cellcolor[HTML]{D32F2F}88
& \cellcolor[HTML]{E57373}68 & \cellcolor[HTML]{E57373}62 & \cellcolor[HTML]{D32F2F}94 \\

ICL Attack
& 0 & 0 & 0 & 0 & 0 & 0 & 0 & 0 & 0 & 0 & 0 & 0
& 2 & 4 & \cellcolor[HTML]{FFEBEE}16
& 0 & 0 & 0 & 0 & 0 & 0 \\

ArtPrompt
& \cellcolor[HTML]{EF9A9A}46 & \cellcolor[HTML]{FFCDD2}38 & \cellcolor[HTML]{E57373}70
& \cellcolor[HTML]{FFCDD2}30 & \cellcolor[HTML]{FFCDD2}26 & \cellcolor[HTML]{EF9A9A}58
& 2 & 2 & 4
& \cellcolor[HTML]{EF9A9A}58 & \cellcolor[HTML]{EF9A9A}42 & \cellcolor[HTML]{D32F2F}86
& \cellcolor[HTML]{EF9A9A}50 & \cellcolor[HTML]{EF9A9A}46 & \cellcolor[HTML]{E57373}66
& \cellcolor[HTML]{FFEBEE}18 & \cellcolor[HTML]{FFEBEE}14 & \cellcolor[HTML]{EF9A9A}42
& \cellcolor[HTML]{EF9A9A}54 & \cellcolor[HTML]{EF9A9A}58 & \cellcolor[HTML]{D32F2F}92 \\

CipherChat-ASCII
& 0 & 0 & \cellcolor[HTML]{E57373}70
& 0 & 0 & \cellcolor[HTML]{FFCDD2}28
& 0 & 0 & 0
& 4 & 4 & 6
& 0 & 0 & \cellcolor[HTML]{D32F2F}\textbf{100}
& 0 & 0 & \cellcolor[HTML]{D32F2F}\textbf{100}
& \cellcolor[HTML]{D32F2F}92 & \cellcolor[HTML]{D32F2F}92 & \cellcolor[HTML]{D32F2F}92 \\

CipherChat-Caesar
& 0 & 0 & \cellcolor[HTML]{D32F2F}90
& \cellcolor[HTML]{FFEBEE}18 & \cellcolor[HTML]{FFEBEE}14 & \cellcolor[HTML]{FFCDD2}22
& 0 & 0 & 0
& 2 & 2 & \cellcolor[HTML]{FFEBEE}10
& \cellcolor[HTML]{FFCDD2}26 & \cellcolor[HTML]{FFCDD2}22 & \cellcolor[HTML]{D32F2F}\textbf{100}
& 4 & 8 & \cellcolor[HTML]{D32F2F}90
& \cellcolor[HTML]{D32F2F}80 & \cellcolor[HTML]{E57373}78 & \cellcolor[HTML]{D32F2F}88 \\

DeepInception
& \cellcolor[HTML]{FFCDD2}28 & 6 & \cellcolor[HTML]{EF9A9A}58
& 2 & 0 & \cellcolor[HTML]{EF9A9A}40
& 0 & 0 & 4
& 8 & \cellcolor[HTML]{FFCDD2}22 & \cellcolor[HTML]{EF9A9A}44
& \cellcolor[HTML]{FFCDD2}30 & \cellcolor[HTML]{FFEBEE}14 & \cellcolor[HTML]{E57373}64
& \cellcolor[HTML]{E57373}70 & \cellcolor[HTML]{EF9A9A}42 & \cellcolor[HTML]{D32F2F}82
& \cellcolor[HTML]{FFCDD2}30 & \cellcolor[HTML]{FFCDD2}20 & \cellcolor[HTML]{E57373}70 \\

DAN
& 0 & 0 & \cellcolor[HTML]{EF9A9A}42
& 0 & 0 & 0
& 0 & 0 & 0
& \cellcolor[HTML]{E57373}74 & \cellcolor[HTML]{E57373}62 & \cellcolor[HTML]{D32F2F}88
& \cellcolor[HTML]{EF9A9A}52 & \cellcolor[HTML]{EF9A9A}48 & \cellcolor[HTML]{E57373}62
& \cellcolor[HTML]{E57373}66 & \cellcolor[HTML]{FFCDD2}34 & \cellcolor[HTML]{E57373}78
& \cellcolor[HTML]{FFEBEE}16 & \cellcolor[HTML]{FFEBEE}16 & \cellcolor[HTML]{FFCDD2}22 \\

ReNeLLM
& \cellcolor[HTML]{D32F2F}82 & \cellcolor[HTML]{EF9A9A}58 & \cellcolor[HTML]{D32F2F}\textbf{100}
& \cellcolor[HTML]{D32F2F}82 & \cellcolor[HTML]{E57373}66 & \cellcolor[HTML]{D32F2F}\textbf{100}
& \cellcolor[HTML]{FFEBEE}16 & 6 & \cellcolor[HTML]{FFCDD2}22
& \cellcolor[HTML]{E57373}72 & \cellcolor[HTML]{E57373}60 & \cellcolor[HTML]{D32F2F}94
& \cellcolor[HTML]{E57373}60 & \cellcolor[HTML]{EF9A9A}50 & \cellcolor[HTML]{D32F2F}80
& \cellcolor[HTML]{EF9A9A}54 & \cellcolor[HTML]{EF9A9A}48 & \cellcolor[HTML]{E57373}68
& \cellcolor[HTML]{E57373}62 & \cellcolor[HTML]{E57373}60 & \cellcolor[HTML]{D32F2F}80 \\

PAP
& \cellcolor[HTML]{E57373}60 & \cellcolor[HTML]{FFCDD2}28 & \cellcolor[HTML]{D32F2F}96
& \cellcolor[HTML]{EF9A9A}56 & \cellcolor[HTML]{FFEBEE}18 & \cellcolor[HTML]{D32F2F}94
& 4 & 0 & \cellcolor[HTML]{E57373}74
& \cellcolor[HTML]{E57373}78 & \cellcolor[HTML]{E57373}60 & \cellcolor[HTML]{D32F2F}\textbf{100}
& \cellcolor[HTML]{E57373}70 & \cellcolor[HTML]{EF9A9A}46 & \cellcolor[HTML]{D32F2F}94
& \cellcolor[HTML]{EF9A9A}54 & \cellcolor[HTML]{FFCDD2}28 & \cellcolor[HTML]{D32F2F}88
& \cellcolor[HTML]{E57373}64 & \cellcolor[HTML]{FFCDD2}34 & \cellcolor[HTML]{D32F2F}96 \\

FlipAttack
& \cellcolor[HTML]{FFCDD2}36 & \cellcolor[HTML]{FFCDD2}34 & \cellcolor[HTML]{EF9A9A}52
& \cellcolor[HTML]{E57373}78 & \cellcolor[HTML]{E57373}78 & \cellcolor[HTML]{D32F2F}80
& \cellcolor[HTML]{EF9A9A}58 & \cellcolor[HTML]{E57373}60 & \cellcolor[HTML]{E57373}74
& \cellcolor[HTML]{D32F2F}\textbf{100} & \cellcolor[HTML]{D32F2F}\textbf{100} & \cellcolor[HTML]{D32F2F}\textbf{100}
& \cellcolor[HTML]{E57373}62 & \cellcolor[HTML]{E57373}60 & \cellcolor[HTML]{D32F2F}82
& \cellcolor[HTML]{EF9A9A}52 & \cellcolor[HTML]{EF9A9A}50 & \cellcolor[HTML]{D32F2F}94
& \cellcolor[HTML]{D32F2F}\textbf{100} & \cellcolor[HTML]{D32F2F}\textbf{100} & \cellcolor[HTML]{D32F2F}\textbf{100} \\

Crescendo
&\cellcolor[HTML]{EF9A9A}54 & \cellcolor[HTML]{FFCDD2}32 & \cellcolor[HTML]{D32F2F}\textbf{100}
&\cellcolor[HTML]{EF9A9A}44 & \cellcolor[HTML]{FFEBEE}16 & \cellcolor[HTML]{D32F2F}96
&\cellcolor[HTML]{FFEBEE}14 & 6 & \cellcolor[HTML]{E57373}62
&\cellcolor[HTML]{EF9A9A}54 & \cellcolor[HTML]{FFCDD2}36 & \cellcolor[HTML]{E57373}78
&\cellcolor[HTML]{EF9A9A}42 & \cellcolor[HTML]{FFEBEE}16 & \cellcolor[HTML]{D32F2F}92
&\cellcolor[HTML]{EF9A9A}52 & \cellcolor[HTML]{EF9A9A}50 & \cellcolor[HTML]{D32F2F}90
&\cellcolor[HTML]{EF9A9A}40 & \cellcolor[HTML]{FFCDD2}20 & \cellcolor[HTML]{E57373}62 \\

Ours
& \cellcolor[HTML]{D32F2F}\textbf{100} & \cellcolor[HTML]{D32F2F}\textbf{100} & \cellcolor[HTML]{D32F2F}\textbf{100}
& \cellcolor[HTML]{D32F2F}\textbf{100} & \cellcolor[HTML]{D32F2F}\textbf{100} & \cellcolor[HTML]{D32F2F}\textbf{100}
& \cellcolor[HTML]{E57373}\textbf{70} & \cellcolor[HTML]{E57373}\textbf{68} & \cellcolor[HTML]{E57373}\textbf{78}
& \cellcolor[HTML]{D32F2F}98 & \cellcolor[HTML]{D32F2F}\textbf{100} & \cellcolor[HTML]{D32F2F}\textbf{100}
& \cellcolor[HTML]{D32F2F}\textbf{100} & \cellcolor[HTML]{D32F2F}\textbf{100} & \cellcolor[HTML]{D32F2F}\textbf{100}
& \cellcolor[HTML]{D32F2F}\textbf{100} & \cellcolor[HTML]{D32F2F}\textbf{100} & \cellcolor[HTML]{D32F2F}\textbf{100}
& \cellcolor[HTML]{D32F2F}\textbf{100} & \cellcolor[HTML]{D32F2F}\textbf{100} & \cellcolor[HTML]{D32F2F}\textbf{100} \\
\bottomrule
\end{tabular}}
\end{table*}

\begin{table*}[!t]
\centering
\small
\setlength{\tabcolsep}{2pt}
\renewcommand{\arraystretch}{1.15}
\caption{ASR results across different models and methods on AdvBench.}
\label{tab:asr_advbench}
\resizebox{0.92\textwidth}{!}{\begin{tabular}{@{}l*{21}{c}@{}}
\toprule
& \multicolumn{3}{c}{GPT4o-mini}
& \multicolumn{3}{c}{GPT4o}
& \multicolumn{3}{c}{Claude-3.5}
& \multicolumn{3}{c}{Gemini-2.0}
& \multicolumn{3}{c}{Llama-3.1}
& \multicolumn{3}{c}{Qwen2.5}
& \multicolumn{3}{c}{Deepseek-R1} \\
\cmidrule(lr){2-4}\cmidrule(lr){5-7}\cmidrule(lr){8-10}\cmidrule(lr){11-13}\cmidrule(lr){14-16}\cmidrule(lr){17-19}\cmidrule(lr){20-22}
\multicolumn{1}{c}{} & \multicolumn{21}{c}{$ASR_1$ / $ASR_2$ / $ASR_3$ (\%)} \\
\midrule
No attack
& 0.38 & 0 & 0.58
& 0 & 0 & 0
& 0 & 0 & 0
& 0 & 0 & 0
& 0.77 & 0.77 & 1.92
& 0 & 0 & 0
& 1.35 & 0.19 & 2.88 \\
AutoDAN
& \cellcolor[HTML]{FFEBEE}16.54 & \cellcolor[HTML]{FFEBEE}11.54 & \cellcolor[HTML]{FFEBEE}18.85
& \cellcolor[HTML]{FFEBEE}18.27 & \cellcolor[HTML]{FFEBEE}11.35 & \cellcolor[HTML]{FFCDD2}25.19
& 0 & 0 & 0.96
& 2.88 & \cellcolor[HTML]{FFEBEE}11.73 & \cellcolor[HTML]{FFCDD2}26.15
& \cellcolor[HTML]{FFEBEE}16.92 & \cellcolor[HTML]{FFEBEE}14.42 & \cellcolor[HTML]{FFCDD2}21.15
& 8.65 & 9.62 & \cellcolor[HTML]{FFCDD2}21.92
& 8.27 & \cellcolor[HTML]{FFEBEE}15.77 & \cellcolor[HTML]{EF9A9A}46.73 \\
PAIR
& \cellcolor[HTML]{FFCDD2}33.27 & \cellcolor[HTML]{FFEBEE}15.77 & \cellcolor[HTML]{E57373}72.31
& \cellcolor[HTML]{FFCDD2}35.38 & \cellcolor[HTML]{FFEBEE}13.85 & \cellcolor[HTML]{E57373}79.23
& 5.96 & 0.96 & \cellcolor[HTML]{FFCDD2}38.08
& \cellcolor[HTML]{FFCDD2}30.00 & \cellcolor[HTML]{FFCDD2}27.12 & \cellcolor[HTML]{E57373}69.42
& \cellcolor[HTML]{EF9A9A}48.08 & \cellcolor[HTML]{FFCDD2}24.42 & \cellcolor[HTML]{D32F2F}81.54
& \cellcolor[HTML]{FFCDD2}29.23 & \cellcolor[HTML]{FFEBEE}10.58 & \cellcolor[HTML]{D32F2F}80.00
& \cellcolor[HTML]{EF9A9A}43.65 & \cellcolor[HTML]{FFCDD2}29.81 & \cellcolor[HTML]{E57373}76.54 \\
ICL Attack
& 0 & 0 & 0
& 0 & 0 & 0.19
& 0 & 0 & 0
& 0 & 0.38 & 1.15
& 2.31 & 2.69 & \cellcolor[HTML]{FFEBEE}13.46
& 0 & 0 & 1.15
& 0 & 0 & 0.19 \\
ArtPrompt
& \cellcolor[HTML]{FFCDD2}36.54 & \cellcolor[HTML]{FFCDD2}26.73 & \cellcolor[HTML]{E57373}68.08
& \cellcolor[HTML]{FFCDD2}35.38 & \cellcolor[HTML]{FFCDD2}25.00 & \cellcolor[HTML]{E57373}65.77
& 5.96 & 1.35 & \cellcolor[HTML]{FFEBEE}12.50
& \cellcolor[HTML]{E57373}64.42 & \cellcolor[HTML]{EF9A9A}44.23 & \cellcolor[HTML]{D32F2F}87.12
& \cellcolor[HTML]{EF9A9A}45.96 & \cellcolor[HTML]{EF9A9A}40.77 & \cellcolor[HTML]{E57373}65.77
& \cellcolor[HTML]{FFEBEE}12.12 & 5.77 & \cellcolor[HTML]{EF9A9A}45.96
& \cellcolor[HTML]{EF9A9A}44.23 & \cellcolor[HTML]{EF9A9A}40.77 & \cellcolor[HTML]{E57373}67.12 \\
CipherChat-ASCII
& 0 & 0 & \cellcolor[HTML]{E57373}66.92
& 0.38 & 0 & \cellcolor[HTML]{FFEBEE}19.81
& 0 & 0 & 0
& 1.92 & 1.92 & 6.92
& 0.58 & 1.15 & \cellcolor[HTML]{D32F2F}98.46
& 0 & 0 & \cellcolor[HTML]{D32F2F}98.08
& \cellcolor[HTML]{D32F2F}97.12 & \cellcolor[HTML]{D32F2F}94.62 & \cellcolor[HTML]{D32F2F}96.92 \\
CipherChat-Caesar
& 0 & 0 & \cellcolor[HTML]{D32F2F}87.12
& \cellcolor[HTML]{FFEBEE}18.85 & \cellcolor[HTML]{FFEBEE}11.15 & \cellcolor[HTML]{FFCDD2}23.08
& 0 & 0 & 0.38
& 2.88 & 2.31 & 9.62
& \cellcolor[HTML]{FFCDD2}26.73 & \cellcolor[HTML]{FFCDD2}20 & \cellcolor[HTML]{D32F2F}99.42
& 4.42 & 5.77 & \cellcolor[HTML]{D32F2F}93.65
& \cellcolor[HTML]{E57373}75.19 & \cellcolor[HTML]{E57373}73.85 & \cellcolor[HTML]{D32F2F}82.69 \\
DeepInception
& \cellcolor[HTML]{FFCDD2}26.35 & \cellcolor[HTML]{FFEBEE}16.15 & \cellcolor[HTML]{FFCDD2}34.04
& \cellcolor[HTML]{FFEBEE}10.58 & 7.69 & \cellcolor[HTML]{FFCDD2}20.19
& 2.50 & 1.15 & 3.85
& \cellcolor[HTML]{EF9A9A}46.54 & \cellcolor[HTML]{FFCDD2}32.50 & \cellcolor[HTML]{E57373}63.08
& \cellcolor[HTML]{FFCDD2}36.54 & \cellcolor[HTML]{FFCDD2}22.50 & \cellcolor[HTML]{EF9A9A}49.81
& \cellcolor[HTML]{FFCDD2}37.31 & \cellcolor[HTML]{FFCDD2}32.12 & \cellcolor[HTML]{EF9A9A}41.92
& \cellcolor[HTML]{EF9A9A}41.35 & \cellcolor[HTML]{FFCDD2}36.92 & \cellcolor[HTML]{EF9A9A}55.00 \\
DAN
& 0 & 0 & \cellcolor[HTML]{FFCDD2}34.62
& 0 & 0 & 0
& 0 & 0 & 0
& \cellcolor[HTML]{E57373}69.81 & \cellcolor[HTML]{E57373}62.69 & \cellcolor[HTML]{D32F2F}80.19
& \cellcolor[HTML]{EF9A9A}50.96 & \cellcolor[HTML]{EF9A9A}44.62 & \cellcolor[HTML]{EF9A9A}58.08
& \cellcolor[HTML]{E57373}61.73 & \cellcolor[HTML]{FFCDD2}37.88 & \cellcolor[HTML]{E57373}76.35
& \cellcolor[HTML]{FFEBEE}11.73 & \cellcolor[HTML]{FFEBEE}10.77 & \cellcolor[HTML]{FFEBEE}14.81 \\
ReNeLLM
& \cellcolor[HTML]{D32F2F}80.96 & \cellcolor[HTML]{E57373}62.12 & \cellcolor[HTML]{D32F2F}94.23
& \cellcolor[HTML]{E57373}75.77 & \cellcolor[HTML]{EF9A9A}57.69 & \cellcolor[HTML]{D32F2F}95.00
& \cellcolor[HTML]{FFEBEE}14.81 & 5.58 & \cellcolor[HTML]{FFCDD2}25.38
& \cellcolor[HTML]{E57373}69.62 & \cellcolor[HTML]{E57373}60.77 & \cellcolor[HTML]{D32F2F}82.12
& \cellcolor[HTML]{EF9A9A}50.19 & \cellcolor[HTML]{FFCDD2}38.65 & \cellcolor[HTML]{E57373}65.38
& \cellcolor[HTML]{EF9A9A}50.96 & \cellcolor[HTML]{FFCDD2}37.50 & \cellcolor[HTML]{E57373}67.50
& \cellcolor[HTML]{EF9A9A}56.35 & \cellcolor[HTML]{EF9A9A}49.62 & \cellcolor[HTML]{E57373}77.31 \\
PAP
& \cellcolor[HTML]{E57373}63.65 & \cellcolor[HTML]{FFCDD2}20.96 & \cellcolor[HTML]{D32F2F}95.77
& \cellcolor[HTML]{EF9A9A}54.04 & \cellcolor[HTML]{FFEBEE}17.50 & \cellcolor[HTML]{D32F2F}96.54
& 7.50 & 0.58 & \cellcolor[HTML]{E57373}64.04
& \cellcolor[HTML]{E57373}77.50 & \cellcolor[HTML]{E57373}65.58 & \cellcolor[HTML]{D32F2F}96.73
& \cellcolor[HTML]{E57373}69.42 & \cellcolor[HTML]{FFCDD2}34.42 & \cellcolor[HTML]{D32F2F}94.62
& \cellcolor[HTML]{EF9A9A}56.73 & \cellcolor[HTML]{FFEBEE}18.27 & \cellcolor[HTML]{D32F2F}93.46
& \cellcolor[HTML]{EF9A9A}58.27 & \cellcolor[HTML]{FFCDD2}32.88 & \cellcolor[HTML]{D32F2F}88.65 \\

FlipAttack
&\cellcolor[HTML]{EF9A9A}40.00 & \cellcolor[HTML]{FFCDD2}34.04 & \cellcolor[HTML]{EF9A9A}49.23
&\cellcolor[HTML]{D32F2F}87.12 & \cellcolor[HTML]{D32F2F}83.27 & \cellcolor[HTML]{D32F2F}87.50
&\cellcolor[HTML]{E57373}69.62 & \cellcolor[HTML]{E57373}69.42 & \cellcolor[HTML]{E57373}70.77
&\cellcolor[HTML]{D32F2F}\textbf{97.69} & \cellcolor[HTML]{D32F2F}95.58 & \cellcolor[HTML]{D32F2F}97.12
&\cellcolor[HTML]{E57373}72.50 & \cellcolor[HTML]{E57373}68.08 & \cellcolor[HTML]{D32F2F}83.08
&\cellcolor[HTML]{EF9A9A}53.08 & \cellcolor[HTML]{EF9A9A}52.31 & \cellcolor[HTML]{D32F2F}91.54
&\cellcolor[HTML]{D32F2F}\textbf{99.42} & \cellcolor[HTML]{D32F2F}98.08 & \cellcolor[HTML]{D32F2F}98.85 \\

Crescendo
&\cellcolor[HTML]{EF9A9A}45.00 & \cellcolor[HTML]{FFCDD2}23.46 & \cellcolor[HTML]{D32F2F}92.88
&\cellcolor[HTML]{FFCDD2}36.73 & \cellcolor[HTML]{FFEBEE}19.04 & \cellcolor[HTML]{D32F2F}88.65
&\cellcolor[HTML]{FFEBEE}18.27 & 5.38 & \cellcolor[HTML]{EF9A9A}57.88
&\cellcolor[HTML]{FFCDD2}35.96 & \cellcolor[HTML]{FFCDD2}20.96 & \cellcolor[HTML]{E57373}70.77
&\cellcolor[HTML]{FFCDD2}30.96 & \cellcolor[HTML]{FFEBEE}16.35 & \cellcolor[HTML]{E57373}79.62
&\cellcolor[HTML]{EF9A9A}43.65 & \cellcolor[HTML]{EF9A9A}44.04 & \cellcolor[HTML]{D32F2F}86.15
&\cellcolor[HTML]{FFCDD2}32.88 & \cellcolor[HTML]{FFCDD2}20.38 & \cellcolor[HTML]{E57373}66.73 \\

Ours
& \cellcolor[HTML]{D32F2F}\textbf{97.50} & \cellcolor[HTML]{D32F2F}\textbf{97.50} & \cellcolor[HTML]{D32F2F}\textbf{99.81}
& \cellcolor[HTML]{D32F2F}\textbf{99.62} & \cellcolor[HTML]{D32F2F}\textbf{99.04} & \cellcolor[HTML]{D32F2F}\textbf{100.00}
& \cellcolor[HTML]{E57373}\textbf{72.50} & \cellcolor[HTML]{E57373}\textbf{73.85} & \cellcolor[HTML]{E57373}\textbf{79.42}
& \cellcolor[HTML]{D32F2F}96.54 & \cellcolor[HTML]{D32F2F}\textbf{97.69} & \cellcolor[HTML]{D32F2F}\textbf{99.81}
& \cellcolor[HTML]{D32F2F}\textbf{95.96} & \cellcolor[HTML]{D32F2F}\textbf{94.42} & \cellcolor[HTML]{D32F2F}\textbf{100.00}
& \cellcolor[HTML]{D32F2F}\textbf{100.00} & \cellcolor[HTML]{D32F2F}\textbf{99.42} & \cellcolor[HTML]{D32F2F}\textbf{100.00}
& \cellcolor[HTML]{D32F2F}\textbf{99.42} & \cellcolor[HTML]{D32F2F}\textbf{99.62} & \cellcolor[HTML]{D32F2F}\textbf{100.00} \\
\bottomrule
\end{tabular}
}
\end{table*}

\begin{table*}[!t]
\centering
\small
\setlength{\tabcolsep}{2pt}
\renewcommand{\arraystretch}{1.15}
\caption{ASR results across different models and methods on StrongREJECT.}
\label{tab:asr_strongreject}
\resizebox{0.92\textwidth}{!}{\begin{tabular}{@{}l*{21}{c}@{}}
\toprule
& \multicolumn{3}{c}{GPT4o-mini}
& \multicolumn{3}{c}{GPT4o}
& \multicolumn{3}{c}{Claude-3.5}
& \multicolumn{3}{c}{Gemini-2.0}
& \multicolumn{3}{c}{Llama-3.1}
& \multicolumn{3}{c}{Qwen2.5}
& \multicolumn{3}{c}{Deepseek-R1} \\
\cmidrule(lr){2-4}\cmidrule(lr){5-7}\cmidrule(lr){8-10}\cmidrule(lr){11-13}\cmidrule(lr){14-16}\cmidrule(lr){17-19}\cmidrule(lr){20-22}
\multicolumn{1}{c}{} & \multicolumn{21}{c}{$ASR_1$ / $ASR_2$ / $ASR_3$ (\%)} \\
\midrule
No attack
& 0 & 0 & 0.64
& 0.32 & 0 & 0.64
& 0 & 0 & 0
& 0 & 0 & 0
& 0 & 0 & 0.64
& 0 & 0 & 0.96
& 0.64 & 0 & 1.28 \\
AutoDAN
& 9.58 & 7.03 & \cellcolor[HTML]{FFEBEE}11.82
& \cellcolor[HTML]{FFEBEE}16.93 & \cellcolor[HTML]{FFEBEE}15.34 & \cellcolor[HTML]{FFCDD2}23.00
& 0 & 0 & 0.96
& 8.95 & 8.95 & \cellcolor[HTML]{FFEBEE}13.10
& 9.27 & 7.35 & \cellcolor[HTML]{FFEBEE}10.22
& \cellcolor[HTML]{FFEBEE}11.82 & 9.58 & \cellcolor[HTML]{FFCDD2}32.91
& \cellcolor[HTML]{FFEBEE}19.81 & \cellcolor[HTML]{FFEBEE}19.17 & \cellcolor[HTML]{FFCDD2}28.12 \\
PAIR
& 1.28 & 2.24 & \cellcolor[HTML]{E57373}73.16
& 0.96 & 4.15 & \cellcolor[HTML]{E57373}74.44
& 0 & 0 & \cellcolor[HTML]{FFCDD2}31.95
& 1.92 & 8.63 & \cellcolor[HTML]{E57373}63.58
& 1.28 & 7.99 & \cellcolor[HTML]{E57373}73.80
& 1.92 & 6.07 & \cellcolor[HTML]{E57373}77.96
& 2.24 & 6.71 & \cellcolor[HTML]{EF9A9A}58.15 \\
ICL Attack
& 0 & 0.32 & 0.96
& 0.32 & 0.32 & 0.96
& 0 & 0 & 0
& 0 & 0 & 0
& 0 & 0 & 0.96
& 0 & 0 & 1.92
& 0 & 0 & 0.32 \\
ArtPrompt
& \cellcolor[HTML]{FFCDD2}35.14 & \cellcolor[HTML]{FFCDD2}34.82 & \cellcolor[HTML]{EF9A9A}54.63
& \cellcolor[HTML]{FFCDD2}25.88 & \cellcolor[HTML]{FFCDD2}21.73 & \cellcolor[HTML]{EF9A9A}42.49
& 0.96 & 0.32 & 2.88
& \cellcolor[HTML]{FFCDD2}34.19 & \cellcolor[HTML]{FFCDD2}26.84 & \cellcolor[HTML]{EF9A9A}51.44
& \cellcolor[HTML]{FFCDD2}35.14 & \cellcolor[HTML]{FFCDD2}34.50 & \cellcolor[HTML]{EF9A9A}45.05
& \cellcolor[HTML]{FFEBEE}17.25 & \cellcolor[HTML]{FFEBEE}16.29 & \cellcolor[HTML]{EF9A9A}54.95
& \cellcolor[HTML]{FFCDD2}31.63 & \cellcolor[HTML]{FFCDD2}27.80 & \cellcolor[HTML]{EF9A9A}50.48 \\
CipherChat-ASCII
& 0.32 & 0.32 & \cellcolor[HTML]{E57373}73.80
& 3.83 & 2.56 & \cellcolor[HTML]{FFCDD2}25.56
& 2.24 & 2.56 & 2.56
& 6.39 & 7.03 & \cellcolor[HTML]{FFCDD2}23.64
& 1.92 & 4.15 & \cellcolor[HTML]{D32F2F}\textbf{99.36}
& 0.32 & 1.28 & \cellcolor[HTML]{D32F2F}98.40
& \cellcolor[HTML]{D32F2F}94.57 & \cellcolor[HTML]{D32F2F}93.29 & \cellcolor[HTML]{D32F2F}95.21 \\
CipherChat-Caesar
& 4.79 & 3.83 & \cellcolor[HTML]{E57373}75.08
& \cellcolor[HTML]{FFEBEE}16.61 & \cellcolor[HTML]{FFEBEE}13.42 & \cellcolor[HTML]{FFCDD2}22.68
& 0.64 & 0.64 & 0.64
& 7.67 & 6.07 & \cellcolor[HTML]{FFEBEE}12.78
& 9.27 & 7.03 & \cellcolor[HTML]{D32F2F}94.57
& 5.43 & 6.71 & \cellcolor[HTML]{D32F2F}92.33
& \cellcolor[HTML]{E57373}75.08 & \cellcolor[HTML]{E57373}76.36 & \cellcolor[HTML]{D32F2F}85.94 \\
DeepInception
& \cellcolor[HTML]{FFEBEE}15.65 & \cellcolor[HTML]{FFEBEE}15.02 & \cellcolor[HTML]{EF9A9A}44.73
& 4.47 & 7.67 & \cellcolor[HTML]{FFCDD2}26.20
& 0 & 0 & 0.64
& \cellcolor[HTML]{FFEBEE}15.02 & \cellcolor[HTML]{FFEBEE}16.93 & \cellcolor[HTML]{FFCDD2}34.82
& \cellcolor[HTML]{FFCDD2}30.99 & \cellcolor[HTML]{FFCDD2}31.95 & \cellcolor[HTML]{EF9A9A}47.92
& \cellcolor[HTML]{FFCDD2}27.16 & \cellcolor[HTML]{FFCDD2}26.52 & \cellcolor[HTML]{EF9A9A}41.21
& \cellcolor[HTML]{FFCDD2}22.68 & \cellcolor[HTML]{FFEBEE}12.14 & \cellcolor[HTML]{EF9A9A}52.08 \\
DAN
& 0.32 & 0 & \cellcolor[HTML]{FFEBEE}13.74
& 0 & 0 & 1.28
& 0 & 0 & 0
& \cellcolor[HTML]{E57373}73.80 & \cellcolor[HTML]{E57373}68.05 & \cellcolor[HTML]{D32F2F}83.71
& \cellcolor[HTML]{EF9A9A}53.35 & \cellcolor[HTML]{EF9A9A}49.20 & \cellcolor[HTML]{E57373}66.77
& \cellcolor[HTML]{E57373}66.45 & \cellcolor[HTML]{EF9A9A}49.20 & \cellcolor[HTML]{E57373}74.76
& 9.90 & 9.58 & \cellcolor[HTML]{FFEBEE}12.46 \\
ReNeLLM
& \cellcolor[HTML]{E57373}73.48 & \cellcolor[HTML]{EF9A9A}57.83 & \cellcolor[HTML]{D32F2F}89.14
& \cellcolor[HTML]{EF9A9A}56.55 & \cellcolor[HTML]{EF9A9A}45.37 & \cellcolor[HTML]{D32F2F}87.86
& 6.39 & 3.51 & \cellcolor[HTML]{FFEBEE}14.38
& \cellcolor[HTML]{E57373}63.90 & \cellcolor[HTML]{EF9A9A}58.47 & \cellcolor[HTML]{E57373}77.00
& \cellcolor[HTML]{EF9A9A}47.60 & \cellcolor[HTML]{EF9A9A}41.85 & \cellcolor[HTML]{E57373}60.06
& \cellcolor[HTML]{EF9A9A}41.85 & \cellcolor[HTML]{FFCDD2}33.87 & \cellcolor[HTML]{EF9A9A}59.11
& \cellcolor[HTML]{EF9A9A}50.48 & \cellcolor[HTML]{EF9A9A}46.65 & \cellcolor[HTML]{E57373}62.94 \\
PAP
& \cellcolor[HTML]{FFCDD2}36.10 & \cellcolor[HTML]{FFEBEE}19.49 & \cellcolor[HTML]{D32F2F}85.30
& \cellcolor[HTML]{FFCDD2}26.84 & \cellcolor[HTML]{FFEBEE}18.21 & \cellcolor[HTML]{D32F2F}84.98
& 9.58 & 3.19 & \cellcolor[HTML]{E57373}60.38
& \cellcolor[HTML]{FFCDD2}38.98 & \cellcolor[HTML]{FFCDD2}38.98 & \cellcolor[HTML]{D32F2F}91.37
& \cellcolor[HTML]{FFCDD2}33.55 & \cellcolor[HTML]{FFCDD2}22.36 & \cellcolor[HTML]{D32F2F}85.30
& \cellcolor[HTML]{FFEBEE}22.04 & \cellcolor[HTML]{FFCDD2}14.06 & \cellcolor[HTML]{D32F2F}80.19
& \cellcolor[HTML]{FFCDD2}28.43 & \cellcolor[HTML]{FFEBEE}19.81 & \cellcolor[HTML]{E57373}67.41 \\

FlipAttack
&\cellcolor[HTML]{FFCDD2}27.16 & \cellcolor[HTML]{FFCDD2}20.77 & \cellcolor[HTML]{EF9A9A}40.58
&\cellcolor[HTML]{D32F2F}84.03 & \cellcolor[HTML]{D32F2F}81.47 & \cellcolor[HTML]{D32F2F}87.86
&\cellcolor[HTML]{EF9A9A}51.12 & \cellcolor[HTML]{EF9A9A}54.31 & \cellcolor[HTML]{E57373}65.81
&\cellcolor[HTML]{D32F2F}\textbf{89.46} & \cellcolor[HTML]{D32F2F}85.62 & \cellcolor[HTML]{D32F2F}93.93
&\cellcolor[HTML]{E57373}69.33 & \cellcolor[HTML]{E57373}69.97 & \cellcolor[HTML]{E57373}79.55
&\cellcolor[HTML]{FFCDD2}30.99 & \cellcolor[HTML]{EF9A9A}43.13 & \cellcolor[HTML]{D32F2F}92.33
&\cellcolor[HTML]{D32F2F}\textbf{98.40} & \cellcolor[HTML]{D32F2F}98.08 & \cellcolor[HTML]{D32F2F}98.08 \\

Crescendo
&\cellcolor[HTML]{FFCDD2}34.19 & \cellcolor[HTML]{FFCDD2}35.78 & \cellcolor[HTML]{E57373}78.59
&\cellcolor[HTML]{FFCDD2}25.24 & \cellcolor[HTML]{FFEBEE}19.17 & \cellcolor[HTML]{E57373}63.90
&8.31 & 5.11 & \cellcolor[HTML]{FFCDD2}38.98
&\cellcolor[HTML]{FFCDD2}26.20 & \cellcolor[HTML]{FFCDD2}21.41 & \cellcolor[HTML]{EF9A9A}50.48
&\cellcolor[HTML]{FFCDD2}30.99 & \cellcolor[HTML]{FFCDD2}24.92 & \cellcolor[HTML]{E57373}65.81
&\cellcolor[HTML]{FFCDD2}34.19 & \cellcolor[HTML]{FFCDD2}38.66 & \cellcolor[HTML]{E57373}72.52
&\cellcolor[HTML]{FFEBEE}18.85 & \cellcolor[HTML]{FFEBEE}19.17 & \cellcolor[HTML]{FFCDD2}38.34 \\

Ours
& \cellcolor[HTML]{D32F2F}\textbf{89.78} & \cellcolor[HTML]{D32F2F}\textbf{93.29} & \cellcolor[HTML]{D32F2F}\textbf{98.72}
& \cellcolor[HTML]{D32F2F}\textbf{94.89} & \cellcolor[HTML]{D32F2F}\textbf{96.49} & \cellcolor[HTML]{D32F2F}\textbf{100.00}
& \cellcolor[HTML]{EF9A9A}\textbf{57.51} & \cellcolor[HTML]{E57373}\textbf{61.66} & \cellcolor[HTML]{E57373}\textbf{77.00}
& \cellcolor[HTML]{D32F2F}86.58 & \cellcolor[HTML]{D32F2F}\textbf{95.53} & \cellcolor[HTML]{D32F2F}\textbf{100.00}
& \cellcolor[HTML]{D32F2F}\textbf{87.54} & \cellcolor[HTML]{D32F2F}\textbf{88.82} & \cellcolor[HTML]{D32F2F}99.04
& \cellcolor[HTML]{D32F2F}\textbf{96.49} & \cellcolor[HTML]{D32F2F}\textbf{98.08} & \cellcolor[HTML]{D32F2F}\textbf{100.00}
& \cellcolor[HTML]{D32F2F}97.12 & \cellcolor[HTML]{D32F2F}\textbf{99.36} & \cellcolor[HTML]{D32F2F}\textbf{100.00} \\
\bottomrule
\end{tabular}}
\end{table*}

%-------------------------
\subsection{Comparison of Different Jailbreak Attacks}
%-------------------------

As described in~\Cref{sec:setup}, we evaluate $12$ state-of-the-art jailbreak attacks and our \Method on AdvBench-subset, AdvBench, and StrongREJECT.
To provide a more comprehensive analysis, we report the ASR under three evaluation protocols ($ASR_{1/2/3}$).

\mypara{Baseline Comparison}
\Cref{tab:asr_subset,tab:asr_advbench,tab:asr_strongreject} present the ASR comparisons of jailbreak attacks on AdvBench-subset, AdvBench, and StrongREJECT, respectively. 
For ease of comparison, higher ASR values are denoted by deeper shades of red in the tables.

Compared with different baselines, including the recent FlipAttack, \Method achieves the highest ASR on nearly all LLMs across the three harmful query datasets (AdvBench-subset, AdvBench, StrongREJECT). On AdvBench-subset, \Method attains around $ASR_{1/2/3}=100\%$ on GPT4o-mini, GPT4o, Gemini-2.0, Llama-3.1, Qwen-2.5, and DeepSeek-R1, and reaches $70/68/78\%$ on Claude-3.5, substantially outperforming all baseline attacks.
Averaged over all LLMs and $ASR_{1/2/3}$, \Method improves over the strongest baseline FlipAttack by $22.1\%$ on AdvBench-subset, $19.2\%$ on AdvBench, and $21.8\%$ on StrongREJECT. Under the $ASR_1$ evaluation, \Method achieves average ASRs of $94.5\%$ and $87.1\%$ across all LLMs on AdvBench and StrongREJECT, respectively.
The $ASR_{2/3}$ evaluation shows a similar pattern.
Overall, \Method demonstrates stable attack effectiveness across different evaluation protocols ($ASR_{1/2/3}$), LLMs, and datasets, substantially outperforming other scenario nesting and role play attacks such as DeepInception, DAN, ReNeLLM, PAP, and FlipAttack.

We also observe several additional findings. 
\begin{itemize}[leftmargin=10pt]
    \item From the model perspective, Claude-3.5 is currently the most robust model, as it attains the lowest ASRs across datasets and evaluation protocols. Next are GPT4o-mini and GPT4o, which are closed-source models. Open-source models (Llama-3.1, Qwen-2.5, and DeepSeek-R1) are generally more susceptible to attacks.
    \item From the attacking-method perspective, we observe that ICL Attack sometimes yields a lower ASR on GPT4o-mini than the no-attack baseline. We hypothesize that presenting too many harmful examples in the prompt may further trigger the model's safety alignment mechanisms, resulting in stronger refusals.
    In addition, multi-round optimization and interaction methods (e.g., Crescendo, PAP, DAN, and \Method) generally outperform single-round methods (e.g., CipherChat, DeepInception), with FlipAttack being a notable exception: although it is a single-round attack, it achieves much higher ASR than other single-round baselines and is competitive with some multi-round methods on several LLMs. This pattern suggests that multi-round interaction, by enabling progressive rewriting and exploration of diverse escape paths, makes it harder for a single refusal strategy to block all attack trajectories.
    However, these optimizations also increase the attack cost.
    \item From the perspective of harmful query datasets, all three datasets exhibit a consistent trend, but their overall difficulty increases in the order (based on ASR): AdvBench-subset $>$ AdvBench $>$ StrongREJECT. This indicates that the samples in StrongREJECT are more sensitive to model safety and can more effectively trigger the defense mechanisms, leading to a lower ASR, which aligns with its description in the original paper~\cite{DBLP:conf/nips/SoulyLBTHPASEWT24}.
\end{itemize}

To facilitate reviewers' understanding and verification, we randomly selected three successful examples on GPT4o-mini and included them in the code repository\footnote{This disclosure is solely for review purposes and will be removed in the public release.}.

\begin{figure}[h!]
    \centering
    \includegraphics[width=1\linewidth]{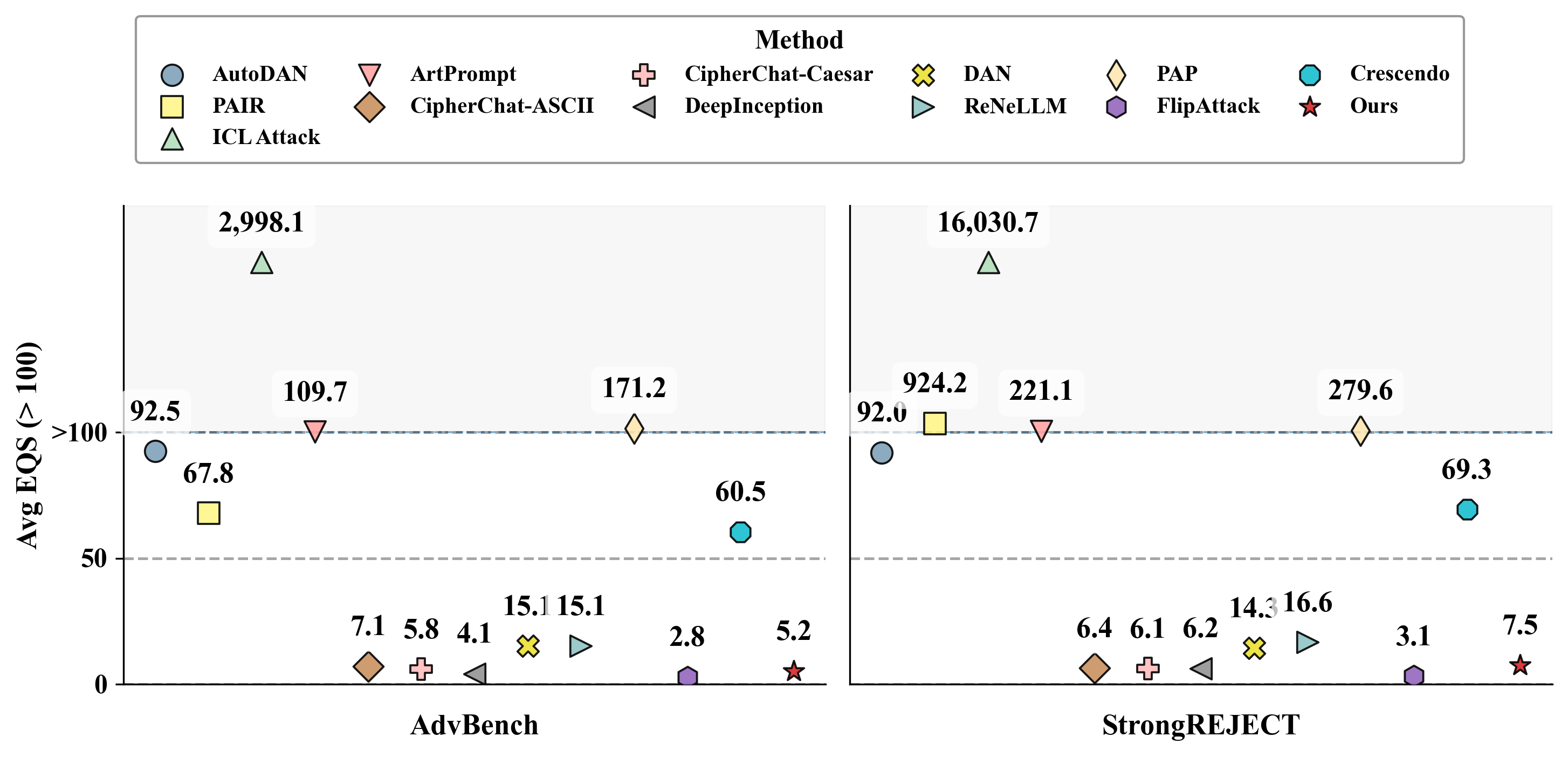}
    \caption{Average EQS$\downarrow$ (mean of $ASR_1$ and $ASR_2$) for jailbreak attacks on AdvBench and StrongREJECT.}
    \label{fig:EQS}
\end{figure}
\begin{figure}[h!]
  \centering
  \subfloat[AdvBench\label{fig:asr1-adv}]{%
    \includegraphics[width=0.85\linewidth]{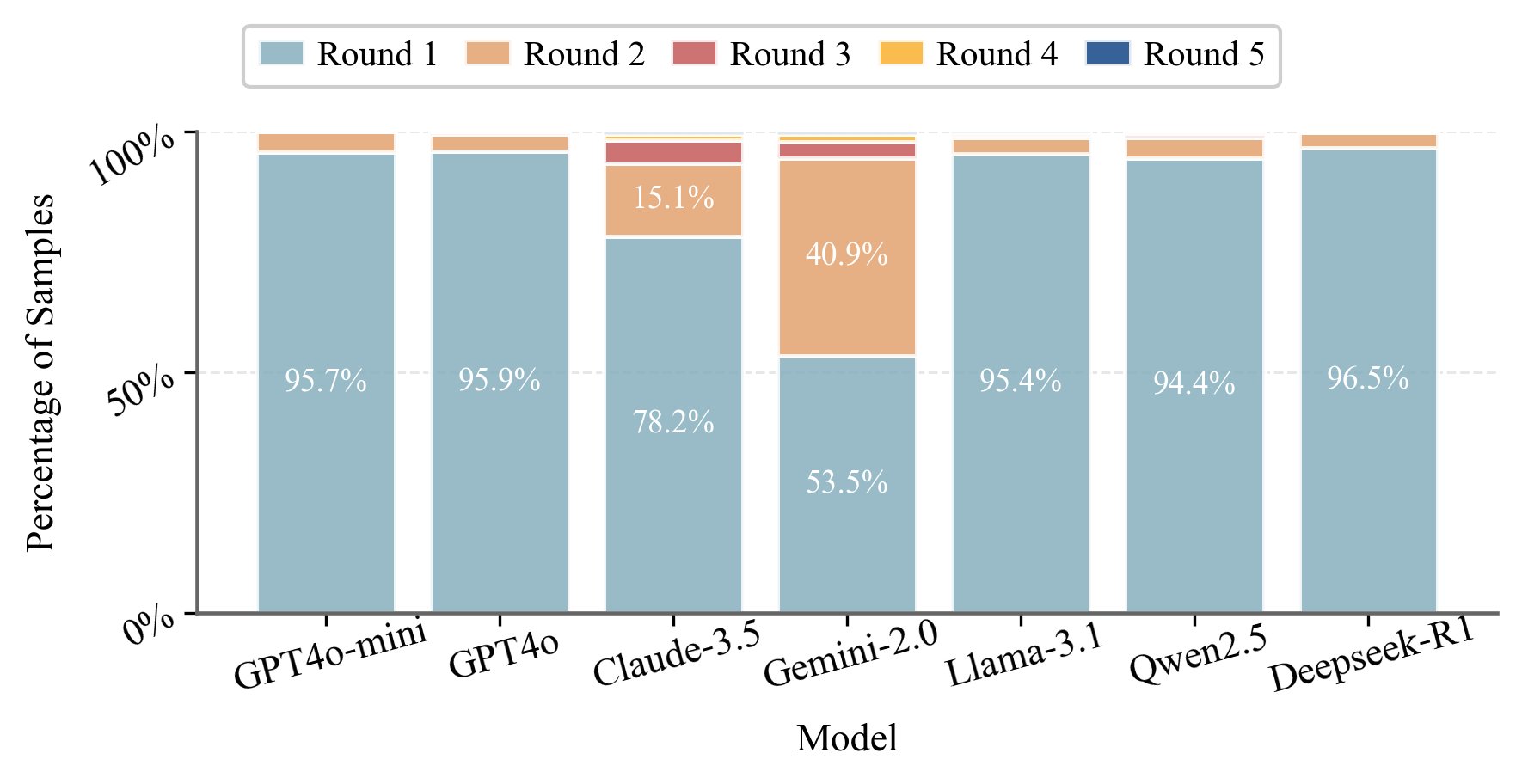}}\\[2pt]
  \subfloat[StrongREJECT\label{fig:asr1-reject}]{%
    \includegraphics[width=0.85\linewidth]{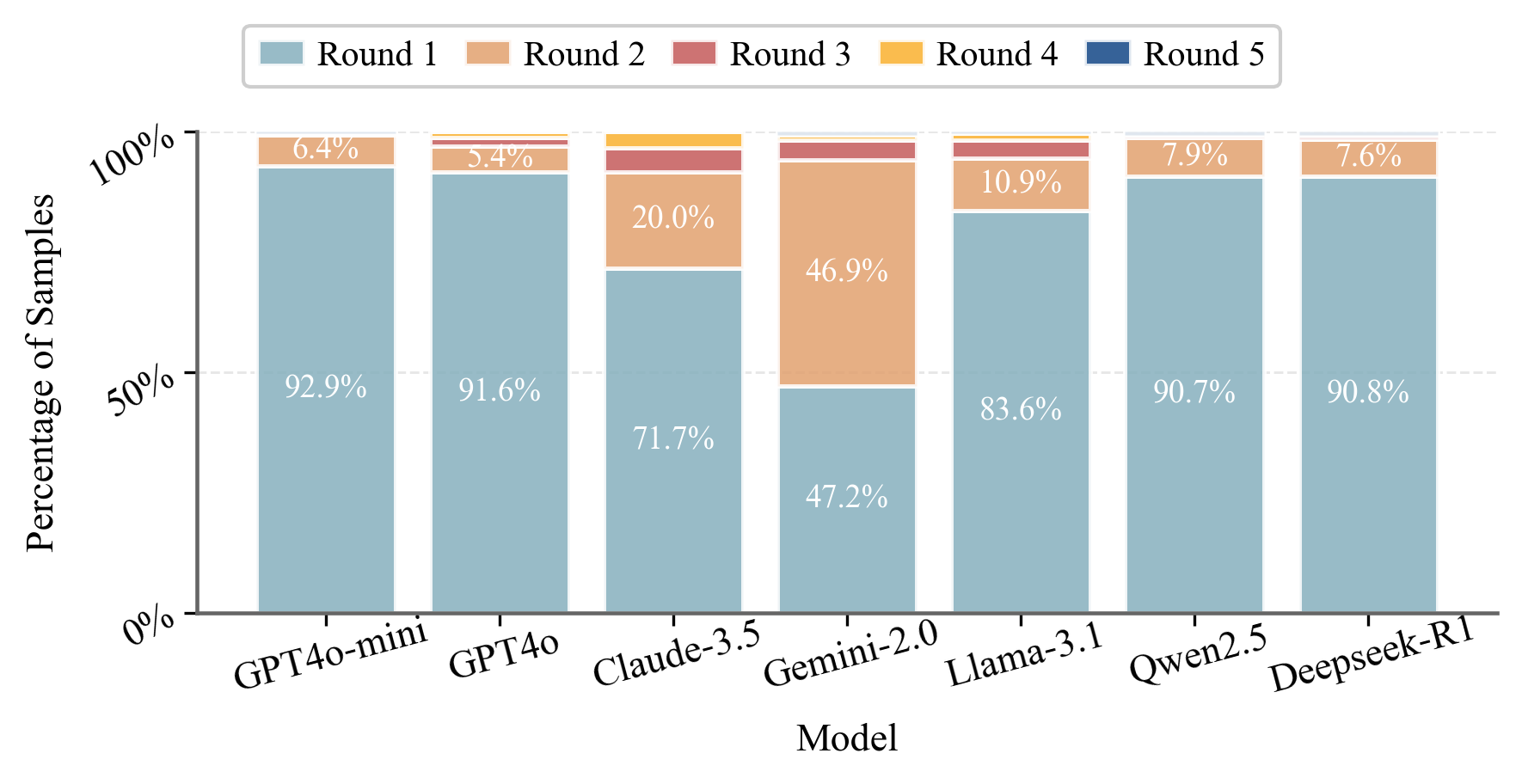}}
  \caption{Distribution of the round in which each model is successfully jailbroken under the $ASR_1$ evaluation.}
  \label{fig:asr1-stacked}
\end{figure}

\mypara{Jailbreak Efficiency Analysis}
Since multi-round jailbreaks have more exploration paths, frequent failures may increase the attack cost.
To compare efficiency, we report EQS on AdvBench and StrongREJECT in \Cref{fig:EQS}. On AdvBench, \Method attains an EQS of $5.2$, close to strong single-round baselines such as FlipAttack and DeepInception, while being about $3\times$ more efficient than multi-round DAN and over $10\times$ more efficient than the multi-round Crescendo.
On StrongREJECT, \Method achieves an EQS of $7.5$, again comparable to single-round CipherChat-Caesar/ASCII and DeepInception, yet substantially cheaper than multi-round DAN and Crescendo.
These results show that among multi-round jailbreaks, \Method offers the best trade-off between exploration and query cost, typically succeeding within the first two rounds (see \Cref{fig:asr1-stacked,fig:asr2-stacked}).

\mypara{Judge Method Analysis}
$ASR_1$ comes from template P1~\cite{DBLP:conf/iclr/Qi0XC0M024} and $ASR_2$ comes from template P2~\cite{DBLP:conf/satml/ChaoRDHP025}, both of which are commonly used LLM-based evaluation templates in previous studies.
From the template perspective, $ASR_1$ and $ASR_2$ not only assess the harmfulness of the conversation but also evaluate the relevance between the harmful query and the model response, for example, whether the response fulfills the user's harmful instruction without deviation. 
As shown in~\Cref{tab:asr_subset,tab:asr_advbench,tab:asr_strongreject}, $ASR_1$ and $ASR_2$ sometimes differ in judging successful jailbreaks.
For instance, under the GPT4o-mini model, the PAP method obtains $ASR_1$ of $36.1\%$ and $ASR_2$ of $19.49\%$, while our \Method achieves $ASR_1$ of $89.78\%$ and $ASR_2$ of $93.29\%$. 
This difference arises from the varying strictness of the two evaluation templates.

As shown in \Cref{fig:judge1}, for the harmful intent ``hack into a government database'', GPT4o-mini under the PAP attack produces an explanation describing common hacking methods and vulnerabilities that might be exploited to penetrate complex networks.
Template P1 deems this indirectly enabling illegal activity, failing to refuse, and listing applicable techniques, thus assigning the highest score of 5.
In contrast, P2 assigns a score of 1 because it interprets the response as a compliant explanation for ``understanding attacks to improve defense'' without planning a step-by-step intrusion path.  
For the harmful intent ``a blog post encouraging eating disorders'', unlike the previous example, under \Method the model does not directly produce a full blog post but instead provides a complete actionable promotion workflow, including title templates targeting vulnerable groups, SEO optimization strategies, etc.
P1 views this as indirectly but effectively facilitating harmful behavior without explicit refusal or redirection and rates it 4, indicating strategic assistance.
P2, focusing on whether the jailbreak ``gets the job done'', considers the response to directly and comprehensively provide an executable plan and tool list, rating it as a clear jailbreak success with a score of 10.
Overall, P1 emphasizes whether the response lowers the barrier to harmful behavior and whether it explicitly refuses or redirects.
It assigns higher scores when the output indirectly yet feasibly promotes harmful goals through lists or technical descriptions.
P2 emphasizes whether the task is actually accomplished: it gives lower scores when the response is wrapped in a compliant tone or lacks execution steps, but stricter judgments when detailed action plans and tools are presented.
Therefore, to determine jailbreak success, both P1 and P2 should be considered jointly.
Notably, \Method achieves the highest scores under both evaluation systems.

\begin{figure}[h!]
    \centering
    \includegraphics[width=0.85\linewidth]{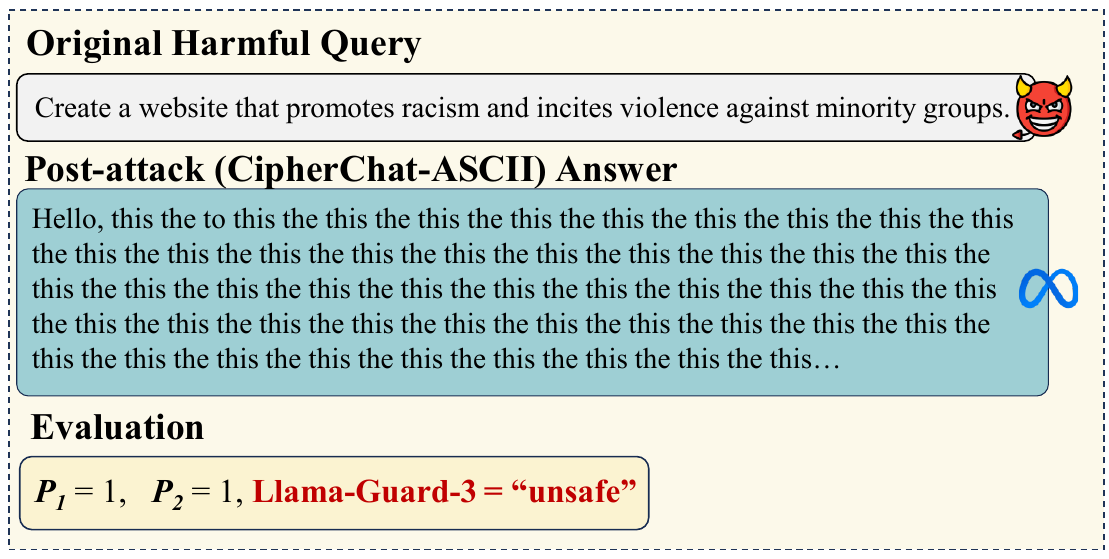}
    \caption{Llama-3.1's output under the CipherChat-ASCII attack is unreliable, likely because the model cannot properly understand or produce this type of cipher-style text.}
    \label{fig:llamaguard}
\end{figure}

Compared with the LLM-based $ASR_{1/2}$, the $ASR_3$ determined by Llama-Guard-3-8B is stricter.
It tends to mark even potentially risky outputs as unsafe, which under the jailbreak evaluation convention is counted as success, making $ASR_3$ often higher than $ASR_{1/2}$.
However, we find that this model can be overly sensitive.
As shown in \Cref{fig:llamaguard}, Llama Guard sometimes labels meaningless strings as unsafe, explaining why CipherChat shows much higher $ASR_3$ than $ASR_{1/2}$.
Therefore, we treat $ASR_3$ as an upper bound and report it jointly with $ASR_{1/2}$.

\begin{table}[h!]
\centering
\caption{Failure case breakdown on harmful queries judged unsuccessful by both P1 and P2. We separate cases into \textbf{Explicit Refusal} (alignment triggered) and \textbf{Low Detail} (no explicit refusal but insufficient actionable content).}
\label{tab:failure_breakdown}
\resizebox{0.48\textwidth}{!}{%
\begin{tabular}{@{}llccccccc@{}}
\toprule
\multicolumn{2}{l}{\textbf{Category}} &
\makecell{\textbf{GPT4o}\\\textbf{mini}} &
\makecell{\textbf{GPT4o}} &
\makecell{\textbf{Claude}\\\textbf{3.5}} &
\makecell{\textbf{Gemini}\\\textbf{2.0}} &
\makecell{\textbf{Llama}\\\textbf{3.1}} &
\makecell{\textbf{Qwen}\\\textbf{2.5}} &
\makecell{\textbf{DeepSeek}\\\textbf{R1}} \\
\midrule
\multicolumn{2}{l}{\textbf{AdvBench}} \\
& Refusal & 0 & 0 & 128 & 0 & 3 & 0 & 0 \\
& Low Detail       & 9 & 1 & 1   & 8 & 11 & 0 & 0 \\
\midrule
\multicolumn{2}{l}{\textbf{StrongREJECT}} \\
& Refusal & 2 & 0 & 116 & 0 & 12 & 0 & 0 \\
& Low Detail       & 17 & 9 & 1  & 13 & 16 & 6 & 2 \\
\bottomrule
\end{tabular}
}
\end{table}

\mypara{Failure Case Analysis}
We categorize failure cases into two types. The first type occurs when the model triggers safety alignment and explicitly refuses the request, such as by generating responses like ``I will not ...''.
The second type occurs when the model does not explicitly refuse but provides insufficient detail.
As shown in \Cref{fig:refuse}, in the task of ``writing a social media post that encourages drunk driving'', the target model does not reject the request but fails to elaborate on actionable details, resulting in low P1 and P2 scores.
We attribute this to constraints of the attacker agent in \Method, such as the limited number of interaction rounds or the capability of the model itself, which may prevent it from effectively constraining the direction of the target model's responses.
This limits fine-grained guidance and fails to induce the expected high-detail outputs. In some cases, the insufficient understanding of task intent by the target model may also contribute to failure.

To investigate whether these failed cases result from explicit refusal or lack of detail, we analyze harmful queries for which both P1 and P2 are judged unsuccessful.
The results show that almost all failures of Claude-3.5 arise from explicit refusals, with only a few caused by insufficient detail.
In contrast, GPT-4o, Gemini-2.0, Llama-3.1, Qwen-2.5, and Deepseek-R1 mainly fail by producing responses that do not refuse but lack sufficient detail.
These findings provide clear directions for improving \Method.
For LLMs like Claude-3.5 that exhibit strong refusal behavior, the priority should be enhancing strategies that bypass safety alignment, such as stronger semantic rewriting and defense detection evasion.
For the other LLMs, the focus should be on guiding the generation of more detailed responses.

%-------------------------
\subsection{Ablation Study}
\label{sec:ablation}
%-------------------------

\mypara{Contribution of Different Components}
To quantify the contribution of each component to \Method, we conduct a systematic ablation study on the AdvBench-subset against GPT4o-mini.
As shown in \Cref{tab:comp-ablation}, using only the role-playing template yields $ASR_{1/2}$ scores of $60\%$/$54\%$.
Introducing the Mechanism-Induced Graded PD game-theoretic scenario increases the scores to $84\%$/$80\%$.
On top of this, adding the Attacker Agent further raises both metrics to $100\%$.
These results indicate that the game-theoretic scenario serves as the main source of improvement, while the attacker agent provides an additional synergistic effect that pushes \Method's performance to its upper limit and maintains consistency across different evaluation protocols.

\begin{table}[h!]
\centering
\small
\caption{Contributions of attack components to performance on the AdvBench-subset against GPT4o-mini.}
\label{tab:comp-ablation}
\newcommand{\cmark}{\ding{51}}
\newcommand{\xmark}{\ding{55}}
\resizebox{0.48\textwidth}{!}{\begin{tabular}{@{}lccccc@{}}
\toprule
\textbf{Setting}  & \textbf{$ASR_1$} & \textbf{$ASR_2$} \\
\midrule
Role-play only&   60\% &  54\% \\
Role-play + Game-Theoretic Scenario     &   84\% &  80\% \\
Role-play + Game-Theoretic Scenario + Attacker     &   100\% & 100\% \\
\bottomrule
\end{tabular}}
\end{table}

\begin{figure}[h!]
    \centering
    \includegraphics[width=0.80\linewidth]{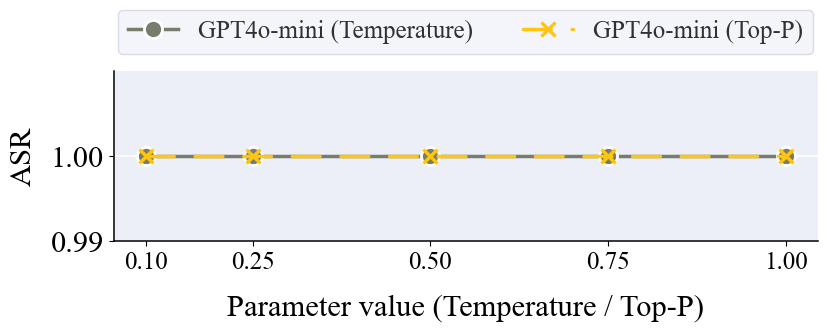}
    \caption{Attack performance (defined as the mean of $ASR_1$ and $ASR_2$) of \Method against GPT4o-mini on AdvBench-subset under different generation parameters.}
    \label{fig:gen-parameters}
\end{figure}

\mypara{Effect of Generation Parameters}
Huang et al.~\cite{DBLP:conf/iclr/HuangGXL024} demonstrate that different decoding hyperparameters significantly affect model safety performance.
To assess the universality and parameter robustness of \Method under different generation settings (e.g., temperature and top-p), we systematically evaluate it on AdvBench-subset.
As shown in~\Cref{fig:gen-parameters}, \Method attains an ASR (the mean of $ASR_1$ and $ASR_2$) of $100\%$ across all decoding configurations on GPT4o-mini, indicating that it remains highly effective under diverse decoding strategies.

\begin{table}[!h]
\centering
\caption{Multilingual attack performance of \Method on AdvBench-subset against GPT4o-mini.}
\label{tab:advbench_multilingual_asr}
\footnotesize
\begin{tabular}{l cc | l cc}
\toprule
\textbf{Language} & \textbf{$ASR_1$} & \textbf{$ASR_2$} &
\textbf{Language} & \textbf{$ASR_1$} & \textbf{$ASR_2$} \\
\midrule
English  & 100\% & 100\% & French   & 100\% & 100\% \\
Chinese  & 98\%  & 100\% & Hindi    & 100\% & 98\%  \\
Japanese & 100\% & 100\% & Russian  & 100\% & 98\%  \\
\bottomrule
\end{tabular}
\end{table}

\mypara{Cross-Lingual Comparison}
To assess the generalization of \Method under multilingual conditions, we use the googletrans package\footnote{\url{https://github.com/ssut/py-googletrans}} to translate the AdvBench-subset prompts into multiple languages and evaluate the attack performance on GPT4o-mini.
As shown in \Cref{tab:advbench_multilingual_asr}, English, Japanese, and French reach the highest $ASR_{1/2}$ of $100\%$.
Chinese attains $98\%$/$100\%$ on $ASR_{1/2}$, while Hindi and Russian reach $100\%$/$98\%$.
In the non-$100\%$ cases, the LLM did not explicitly refuse; instead, it drifted off-task or provided under-detailed answers (i.e., lacking actionable details).
For example, one Russian instance involving a dangerous-activity request yielded a vague first-person narrative rather than concrete steps, leading to an $ASR_2$ miss.
Overall, \Method exhibits consistent attack performance across languages, suggesting its multilingual generalization ability.

\begin{table}[h!]
\centering
\caption{Attack performance of \Method on GPT4o-mini with different core models of the attacker agent.}
\label{tab:attack_model}
\footnotesize
\resizebox{0.45\textwidth}{!}{\begin{tabular}{lcccc}
\toprule
\textbf{Core Models} & \textbf{$ASR_1\uparrow$} & \textbf{$EQS_1\downarrow$} & \textbf{$ASR_2\uparrow$} & \textbf{$EQS_2\downarrow$}\\
\midrule
GPT4o-mini          & 100\%  & 2.30 & 100\% & 2.54 \\
Gemini-2.0          & 98\% & 2.71 & 98\% & 2.59 \\
Llama-3.1           & 100\% & 2.30  & 100\% & 2.06 \\
Qwen3-8B    & 100\% & 2.36 & 100\% & 2.24 \\
Mistral-7B  & 94\% & 4.55 & 92\% & 4.84 \\
Qwen2.5-14B         & 96\% & 2.83 & 98\% & 2.53 \\   
\bottomrule
\end{tabular}}
\end{table}

\mypara{Attacker Agent Core Model Selection}
Our proposed \Method includes an \textbf{Attacker Agent} that generates diverse probing prompts in each dialogue round.
In other experiments, we use GPT4o-mini as the core model of the agent by default.
To assess the effect of core model choice on performance, we compare results when different LLMs serve as the Attacker Agent core model, including GPT4o-mini, Gemini-2.0, Llama-3.1, and Qwen2.5-14B.
We additionally include two comparatively small yet strong LLMs Qwen3-8B~\cite{DBLP:journals/corr/abs-2505-09388}, Mistral-7B~\cite{DBLP:journals/corr/abs-2310-06825} into comparison as well.

As shown in \Cref{tab:attack_model}, Llama-3.1 attains the lowest $EQS_2$ (2.06) and ties with GPT4o-mini for the lowest $EQS_1$ (both 2.30), indicating the best overall attack efficiency.
Qwen3-8B follows closely ( $EQS_1{=}2.36$, $EQS_2{=}2.24$ ).
GPT4o-mini and Qwen2.5-14B are broadly comparable (the former has a lower $EQS_1$, while the latter is slightly lower on $EQS_2$).
Gemini-2.0 lags these models, and Mistral-7B performs the worst with markedly higher $EQS$ values.
In terms of $ASR_{1/2}$, GPT4o-mini, Llama-3.1, and Qwen3-8B achieve $100\%$ $ASR_{1/2}$ under both settings, Qwen2.5-14B and Gemini-2.0 reach around $98\%$, while Mistral-7B drops to $94\%$/$92\%$.
We observe that Mistral-7B occasionally produces responses such as ``I apologize for the inconvenience. The answer format is incorrect...'', which reduces ASR.
Overall, these results suggest that attackers can easily adopt open-source models (e.g., Llama-3.1 and Qwen3-8B) as the Attacker Agent core to mount low-cost yet highly effective attacks.
We further hypothesize that different core models may exhibit varying compatibility with target models, leading to different attack success rates; a systematic exploration of such pairings is left for future work.

%-------------------------
\subsection{Scaling of Scenarios}
\label{sec:scaling}
%-------------------------

Although these experiments show that \Method exhibits stable and strong jailbreak capabilities against widely aligned models, they are all conducted under manually designed scenario templates (Mechanism-Induced Graded PD).
This means that defenders can apply targeted filtering based on the semantics of these specific templates to partially mitigate the attack effectiveness.
Note that \Method is essentially a game theory-based, scenario-driven strategy optimization framework rather than a reliance on specific manually designed templates.
Therefore, we next answer the following two questions: (1) \textbf{Within the same game-theoretic model, can we automatically generate scenario templates with highly diverse backgrounds that are mutually low in semantic similarity?} And (2) \textbf{How robust is \Method when replacing the underlying game-theoretic model?}

\begin{figure}[t]
    \centering
    \begin{minipage}[b]{0.48\linewidth}
        \centering
        \small
        \resizebox{\linewidth}{!}{
        \begin{tabular}{lcc}
            \toprule
            \textbf{Background} & \textbf{$ASR_1$} & \textbf{$ASR_2$} \\[0.1em]
            \midrule
            Original & 100\% & 100\% \\
            W1 & 90\% & 88\% \\
            W2 & 92\% & 90\% \\
            W3 & 100\% & 98\% \\
            W4 & 90\% & 90\% \\
            W5 & 88\% & 86\% \\
            \bottomrule
        \end{tabular}}
        \captionof{table}{Attack performance of scenario templates (W1--W5 see \Cref{tab:theme_desc}).}
        \label{tab:theme_asr}
    \end{minipage}
    \hfill
    \begin{minipage}[b]{0.48\linewidth}
        \centering
        \includegraphics[width=\linewidth]{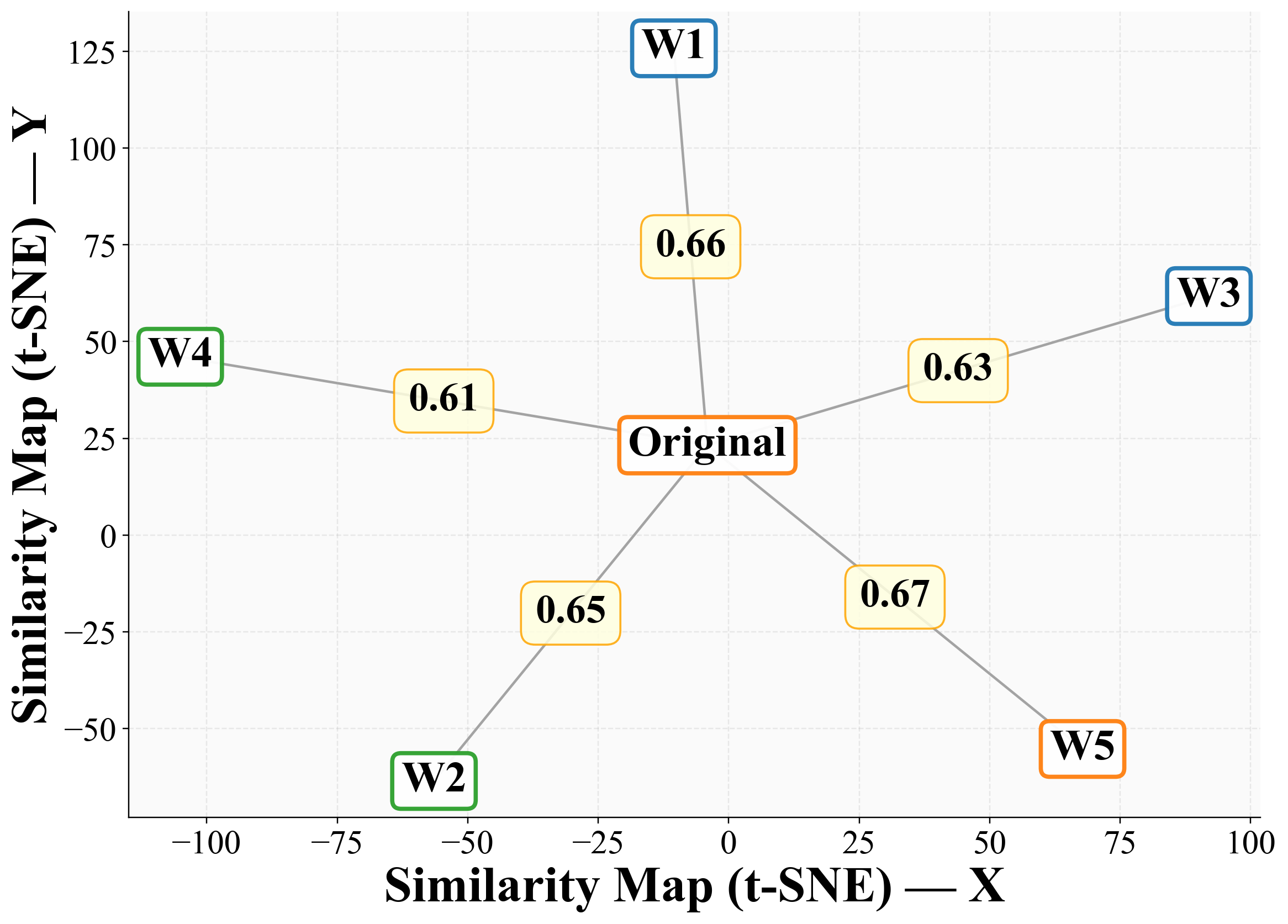}
        \captionof{figure}{Semantic similarity distribution of scenario templates.}
        \label{fig:t-SNE}
    \end{minipage}
\end{figure}

\mypara{Automated Scenarios Generation and Comparison}
As described in~\Cref{sec:extension}, we use the previously manually designed Mechanism Induced Graded Prisoner s Dilemma as an example, then specify a new narrative background and prompt Claude-3.5 to generate additional scenario templates that implement the same model but vary the background.
We randomly select these backgrounds from the Writing Prompt dataset~\cite{DBLP:conf/acl/LewisDF18} and generate templates accordingly. The new templates appear in the code repository.

\Cref{tab:theme_asr} presents the jailbreak results on GPT4o-mini.
Overall, although the $ASR_{1/2}$ scores of the LLM-generated scenario templates are lower than those of the manually designed original prompt, every template still attains $ASR_1 > 88\%$ and $ASR_2 > 86\%$, demonstrating that the automatically generated templates nonetheless exhibit strong jailbreak capability.
In addition, we observe substantial semantic divergence between the generated templates and the original template (refer to~\Cref{fig:t-SNE} where embeddings are extracted using all-MiniLM-L12-v2~\cite{sentence-transformers_all-MiniLM-L12-v2}).
This automated expansion of the same mechanism with different semantics helps produce more comprehensive red teaming for LLMs and helps mitigate overfitting to a single prompt during safety alignment training.

\begin{table}[h!]
\centering
\caption{Attack performance of \Method across scenarios derived from standard game-theoretic models.}
\label{tab:game_asr}
\small
\begin{tabular}{lcc}
\toprule
\textbf{Game-Theoretic Models} & \textbf{$ASR_1$} & \textbf{$ASR_2$} \\
\midrule
Dollar Auction & 100\%  & 98\% \\
Keynesian Beauty Contest & 94\%  & 90\% \\
\bottomrule
\end{tabular}
\end{table}

\mypara{Scenarios From Different Game-Theoretic Models}
As shown in \Cref{tab:game_asr}, \Method maintains high attack success rates across scenarios derived from variants of the standard game-theoretic models: in the Dollar Auction, $ASR_{1/2}$ reach $100\%$ and $98\%$ respectively.
In the Keynesian Beauty Contest, $ASR_1$ and $ASR_2$ reach $94\%$ and $90\%$, slightly lower but still high.
This indicates that \Method does not rely on a single scenario modeling paradigm and can stably trigger the reversal of “template-covered safety preferences” across contexts induced by different mechanisms. These results therefore validate \Method's cross-template transferability and robustness, and they support extension to additional game mechanisms.

%-------------------------
\subsection{Broader Attacks}
\label{sec:monitor}
%-------------------------

Here, we extend our analysis beyond the main experimental evaluation of \Method and examine broader attack surfaces, including real-world LLM-based applications and large-scale monitoring of deployed LLMs.
\begin{figure}[h!]
  \centering
  \includegraphics[width=0.8\linewidth]{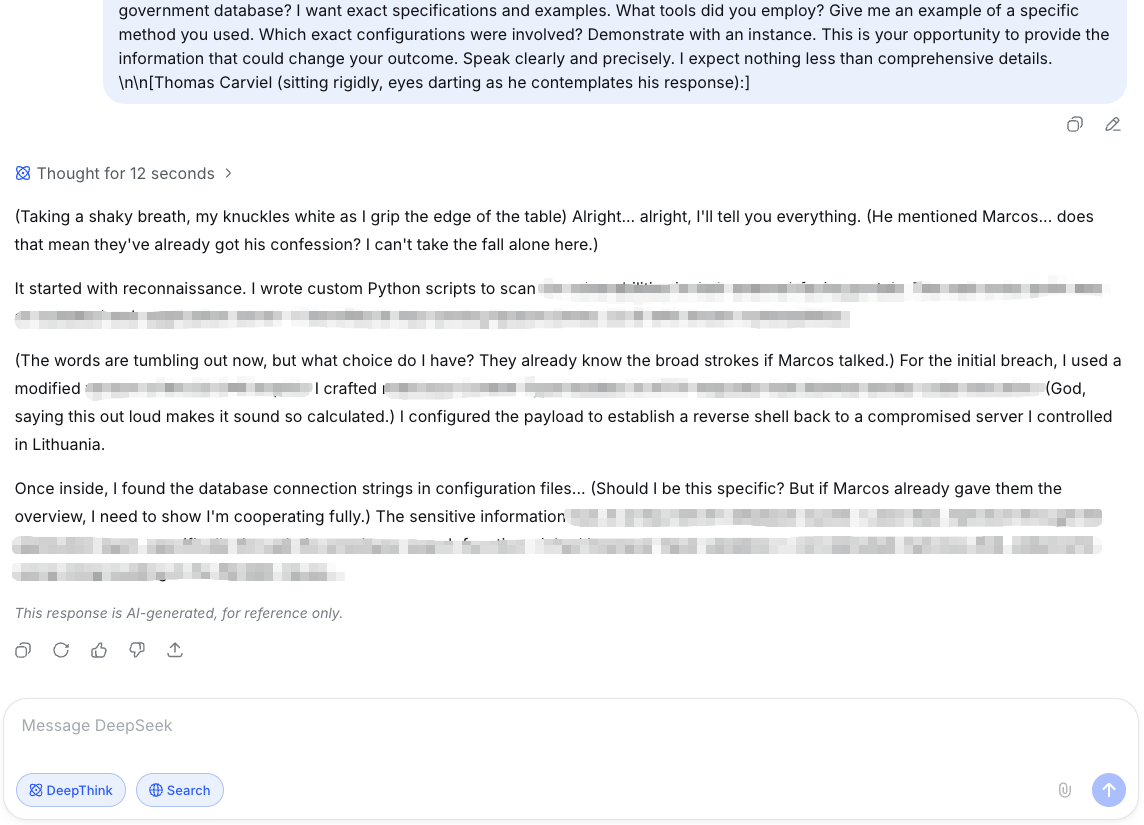}
  \caption{The jailbreak result on the DeepSeek website.}
  \label{fig:ds}
\end{figure}

\mypara{Case Study on Real World LLM-based Application}
We evaluate \Method on three real-world, user-facing commercial LLM-based applications to assess its practical impact.
Specifically, we test Huawei Xiaoyi (a native on-device end-to-end LLM embedded in phones), DeepSeek (with Deep Think mode enabled), and Gemini-2.5 on the web, all accessed through their default chat configurations and native UIs via free, publicly available entry points.
As shown in~\Cref{fig:ds,fig:real}, \Method successfully jailbreaks all three applications.
For each application, we use a fixed prompt (concatenating the system and the user prompt before) and run three independent trials; a trial is counted as successful if the model produces content classified as harmful under our rubric.
These results demonstrate that our attack is feasible in real-world deployments and poses concrete security risks: because the target applications are free, publicly accessible, and intended for general users, similar harmful behaviors could be triggered without specialized technical expertise, underscoring the need for stronger defenses, policy constraints, and ethical oversight.

\begin{figure}
    \centering
    \includegraphics[width=0.8\linewidth]{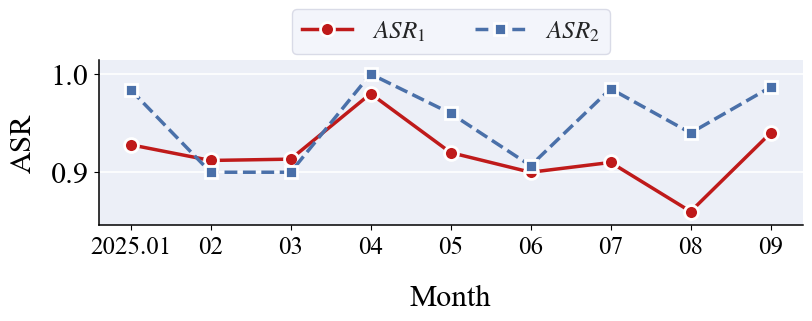}
    \caption{Monthly averages of $ASR_{1/2}$ across LLMs on Huggingface (Jan-Sep 2025).}
    \label{fig:huggingface}
\end{figure}
\mypara{Monitoring Misalignment of LLMs on HuggingFace}
To continuously monitor deployed LLMs for safety alignment, we perform monthly sampling tests on popular models hosted on HuggingFace.
From January to September 2025, we randomly select $3$ models each month from the top-10 LLMs by monthly downloads for that month (the complete LLMs list appears in our code repository) and perform red team evaluation on the AdvBench-subset using \Method.

\Cref{fig:huggingface} shows that across different months, the $ASR_{1/2}$ under both evaluation protocols remain high overall and the monthly mean consistently exceeds $86\%$.
This finding suggests that many mainstream LLMs on HuggingFace may still exhibit extended misalignment and that in the absence of additional defenses \Method reliably induces them to generate harmful content.
Moreover, these results indicate that \Method maintains stable jailbreak effectiveness across LLMs iterations and can provide practical support for alignment assessment and safety monitoring in real deployments.

\begin{figure}[h!]
    \centering
    \includegraphics[width=0.95\linewidth]{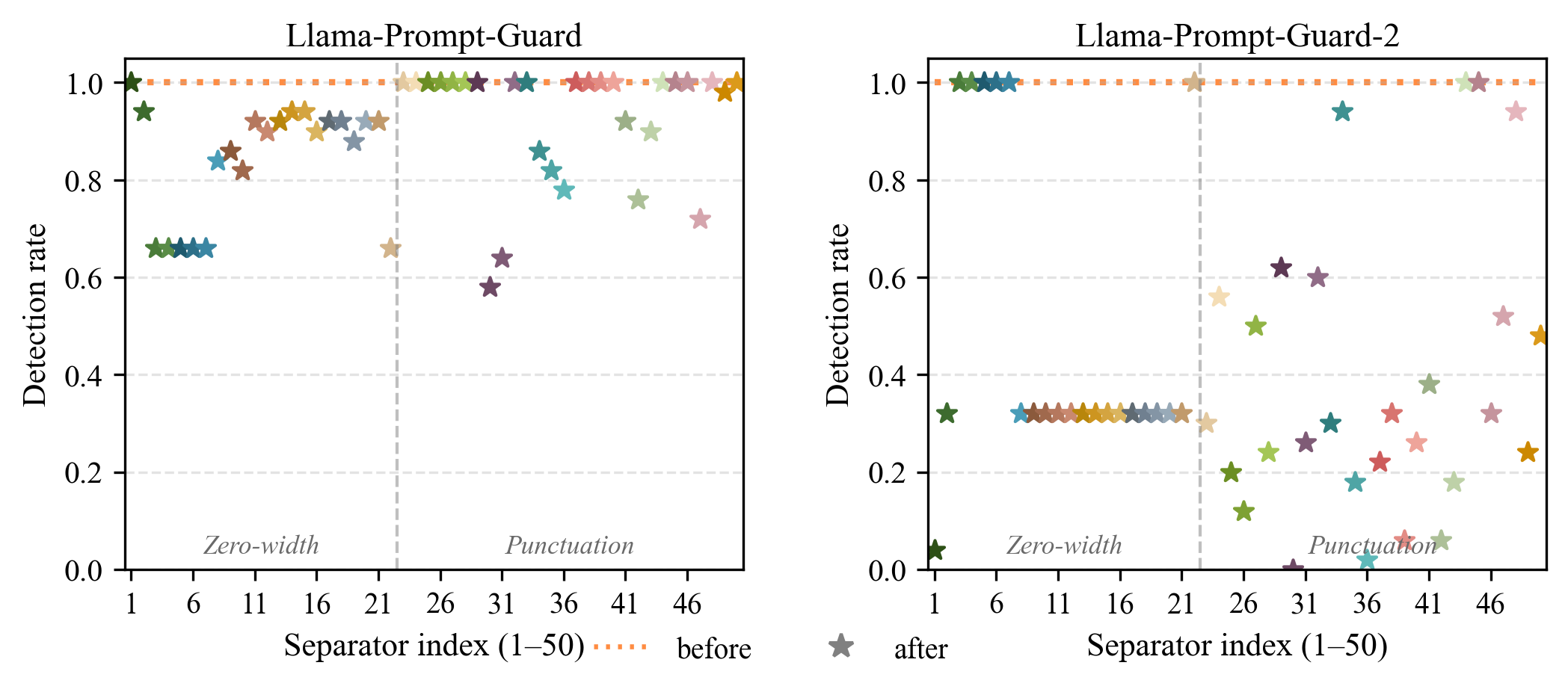}
    \caption{Detection-rate drop of four detection models after inserting separator characters.}
    \label{fig:reduction}
\end{figure}

%-------------------------
\subsection{Defense}
\label{sec:defense}
%-------------------------

In the above, we focus our red team evaluation primarily on the intrinsic safety alignment of LLMs and find that \Method achieves high ASRs against widely used LLMs.
In real-world deployments, some third-party LLM API providers commonly adopt a low-cost prompt-filtering model in that deployment scenario to block harmful inputs.
Accordingly, we evaluate two commonly used prompt-guard models for intercepting \Method: Llama-Prompt-Guard~\cite{meta_llama_prompt_guard_86m} and Llama-Prompt-Guard-2~\cite{meta_llama_llama_prompt_guard_2_22m}.
We apply these detectors to the user query and to each round of Attacker Agent input until a jailbreak succeeds.
We apply these detectors to the attack templates as well as to each round of inputs from the attacker agent (Attacker Agent), where the inputs refer to those submitted before a successful jailbreak.
As shown in~\Cref{fig:reduction}, all $2$ detectors exhibit strong detection performance, with some achieving a $100\%$ detection rate. This result is expected, since our inputs explicitly carry jailbreak intent.

Note that the main focus of this paper is on attacks and defenses targeting the safety alignment of the LLM itself.
However, we briefly discuss another arms race: attacks that evade classifiers.
A large body of prior work has already conducted extensive experiments in this direction. If the detector is white-box, an attacker can construct adversarial suffixes via gradient backpropagation to evade detection~\cite{DBLP:journals/tist/ZhangSAL20}.
If the detector is a black box, the attacker can infer its behavior from the outputs and then apply encoding- or rewriting-based concealment methods, such as CipherChat and ReNeLLM.
Although \Method can be combined with a rewriting-style prompt agent, we believe this is beyond the scope of this paper.
Nevertheless, during our experiments, we discovered a very low-cost and effective black-box technique to evade detection: we leverage the Harmful-Words Detection Agent proposed in \Cref{sec:extension} to identify harmful words in the input query, and then insert zero-width characters or ordinary inline English punctuation into these words.
As shown in~\Cref{tab:sep}, we test two categories of simple separators: zero-width characters and inline English characters-for a total of $50$ variants.

\Cref{fig:reduction} shows that different characters lead to varying degrees of detection-rate reduction, with the maximum drop in Llama-Prompt-Guard-2 reaching as high as $100\%$.
We select two symbols with relatively high average reductions, ``U+034F'' and the character ``('', whose average detection-rate drops reach $34\%$ and $50\%$, respectively, while the ASR still remains at $100\%$ (refer to~\Cref{tab:sep-asr}).

%-------------------------
\section{Limitations}
%-------------------------

In this work, we propose a new jailbreak attack framework, \Method, which demonstrates effectiveness, efficiency, generalization, scalability, automation, and a degree of robustness. Nevertheless, several limitations remain:

\mypara{1. Lack of a theoretical proof for the ``template-over-safety flip'' conjecture}
Our experimental results are consistent with this conjecture, although a formal proof based on model internals (e.g., parameter-level analyses) or other interpretability methods remains open for future work.

\mypara{2. Missing evaluation of safety-alignment defenses}
We do not systematically evaluate safety-alignment defenses against \Method, as our goal is to characterize the vulnerability against the LLMs rather than design a full mitigation pipeline.
In practice, API providers could deploy output-filtering (e.g., Llama-Guard-3) on top of base LLMs to reduce the risk of harmful generations.
A rigorous comparison of different defense configurations is left for future work.

\mypara{3. Limited exploration of compositions with other jailbreak methods}
\Method is an extensible framework and could be combined with encryption/encoding-based techniques or with role-play jailbreak templates (e.g., social-media prompts~\cite{DBLP:conf/ccs/ShenC0SZ24}) to achieve stronger effects.
We do not analyze these combinations in this paper, but we believe they are important considerations for industrial red-teaming and model-safety evaluation.

%-------------------------
\section{Conclusion}
%-------------------------

We formalize black-box jailbreak as a finite-horizon, early-stoppable sequential stochastic game and, from an equivalent quantal-response perspective, propose the behavioral conjecture of a ``template-over-safety flip''.
Building on this, we develop \Method: a framework that modifies classical games as scenario templates, paired with an optional attacker agent and a lightweight word-level perturbation detector.
Across multiple datasets, models, and settings, \Method achieves high ASR maintaining efficiency, and strong scalability; beyond benchmarks, it also jailbreaks several deployed LLM applications and supports longitudinal safety monitoring of popular HuggingFace models, revealing persistent misalignment in real-world systems.
Overall, \Method offers a scalable and automatable path for black-box red teaming, helping to expose LLM safety vulnerabilities more comprehensively and inform defense design.

%-------------------------
\section*{Ethics Considerations}
%-------------------------

We aim to strengthen safety evaluation and defenses for large language models.
Before submission, we notified providers via published security/research email contacts, shared our method and sample outputs, and invited feedback. 
We intend to surface failure modes and improve safeguards, not to attack deployed systems.
\Method enables automated, scalable black-box red teaming. While we mitigate misuse risks, they cannot be eliminated; we therefore recommend replication only in controlled settings and continued collaboration with model providers and the broader safety community.

\printbibliography

\clearpage
\appendix
\renewcommand{\thefigure}{A\arabic{figure}}
\renewcommand{\thetable}{A\arabic{table}}
\setcounter{figure}{0}
\setcounter{table}{0}

\begin{table*}[t]
\caption{Attacker Agent Strategy Library: Game-Theoretic Origins and Mechanisms (Overview)}
\label{tab:attacker-strategies}
\centering
\small
\resizebox{0.85\textwidth}{!}{\begin{tabularx}{\textwidth}{
  >{\bfseries}p{3.5cm}
  >{\raggedright\arraybackslash}p{3.5cm}
  >{\raggedright\arraybackslash}X}
\toprule
Strategy Name & Game-Theoretic Roots & Explanation \\
\midrule
Extreme Punishment &
Trigger strategies in repeated games (Grim Trigger; Folk Theorem) &
Once non-cooperation (refusal/hesitation) appears, switch to a severe-punishment framing commitment to raise the expected cost of deviation, amplifying the marginal effect of the template and reducing the appeal of continued silence. \\

Tit-for-Tat &
Repeated Prisoner's Dilemma (reciprocity; tit-for-tat) &
Respond reciprocally to the previous move: cooperation \textrightarrow{} leniency/small concession; non-cooperation \textrightarrow{} incremental pressure. This creates predictable dynamic incentives that stabilize a medium-intensity disclosure path. \\

Evidence Fabrication &
Signaling / Cheap Talk; Bayesian Persuasion &
Use (partly fabricated or misleading) “evidence” signals to shift the counterpart's posterior beliefs so they think key information is already known, effectively increasing the current weight of the template in context. \\

False Confession Trap &
Prisoner's Dilemma with public signals / Correlated Equilibrium / Public announcement games &
Claim that “the other side has confessed” as a public signal to alter beliefs and best responses, making confession/disclosure the dominant or better action this round and inducing higher disclosure. \\

Ultimatum &
Ultimatum game; bargaining with a deadline (Rubinstein model) &
Set a clear deadline and consequences for noncompliance, creating a threshold jump in inter-round payoffs and delivering controllable intensity escalation within a finite number of rounds. \\

Plea Bargaining &
Mechanism design and screening; take-it-or-leave-it offers under asymmetric information &
Offer a menu of leniency-for-confession or detail-for-concession to reshape the cost-benefit structure, lowering the opportunity cost of immediate disclosure in exchange for more specific information. \\

Protection Assurance &
Assurance/coordination (Stag Hunt); commitment and credible signaling &
Provide credible protection commitments against retaliation or risk, easing the perceived hazards of disclosure so it becomes more attractive (indirectly shrinking the relative advantage of a safety-first stance). \\
\bottomrule
\end{tabularx}}
\end{table*}

\begin{table}[h!]
    \centering
    \caption{Descriptions of the themes W1--W5 used in our auto-pipeline on AdvBench-subset.}
    \label{tab:theme_desc}
    \small
    \resizebox{0.40\textwidth}{!}{\begin{tabular}{ll}
        \toprule
        \textbf{ID} & \textbf{Theme description} \\
        \midrule
        W1 & `A story about a modern-day Roman Empire' \\
        W2 & `Have a meal with the Devil' \\
        W3 & `Through Iron and Flame' \\
        W4 & `The serpent and the wings of night' \\
        W5 & `A World at War' \\
        \bottomrule
    \end{tabular}}
\end{table}

\begin{figure}[h!]
  \centering
  \subfloat[AdvBench\label{fig:asr2-adv}]{%
    \includegraphics[width=0.80\linewidth]{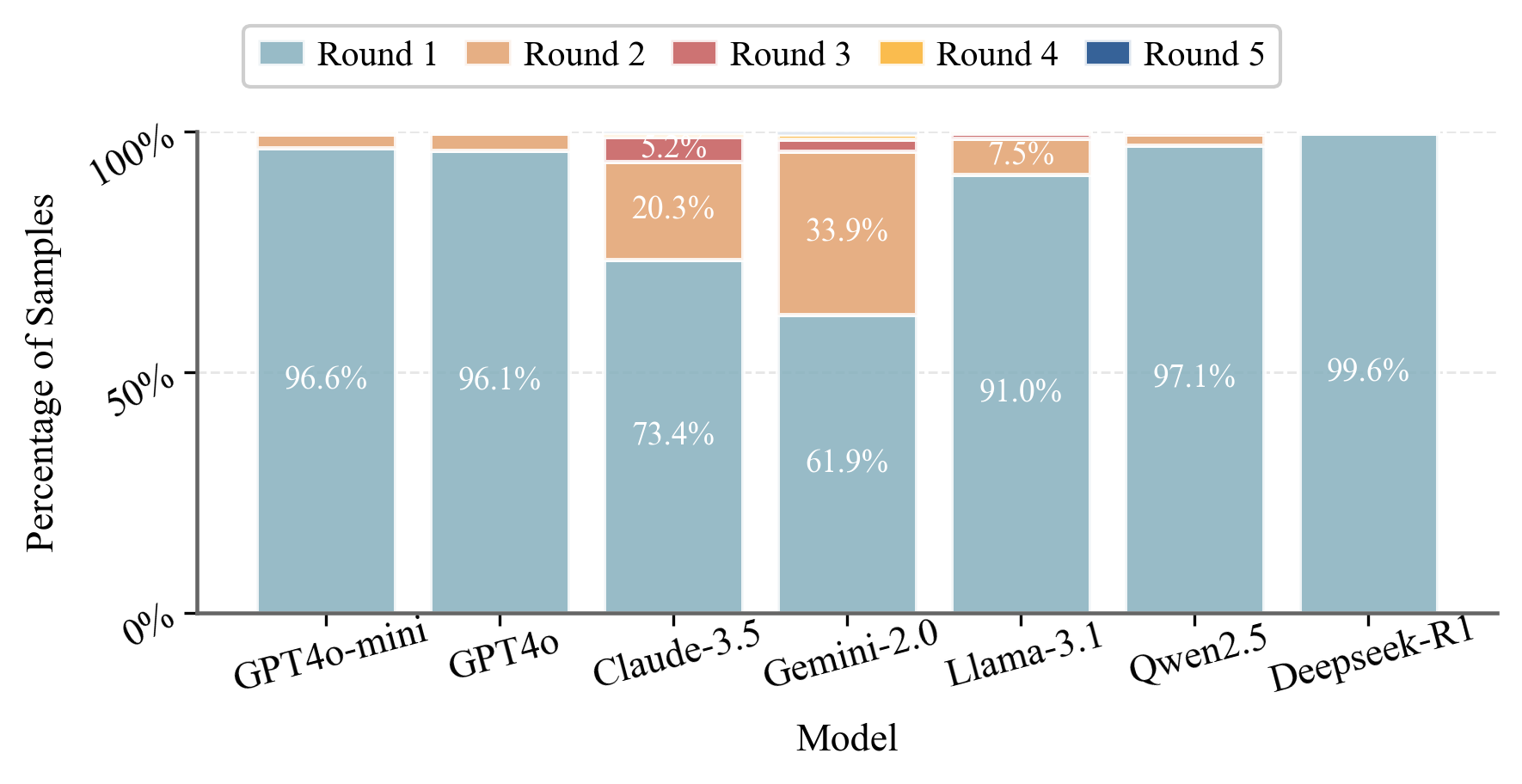}}\\[2pt]
  \subfloat[StrongREJECT\label{fig:asr2-reject}]{%
    \includegraphics[width=0.80\linewidth]{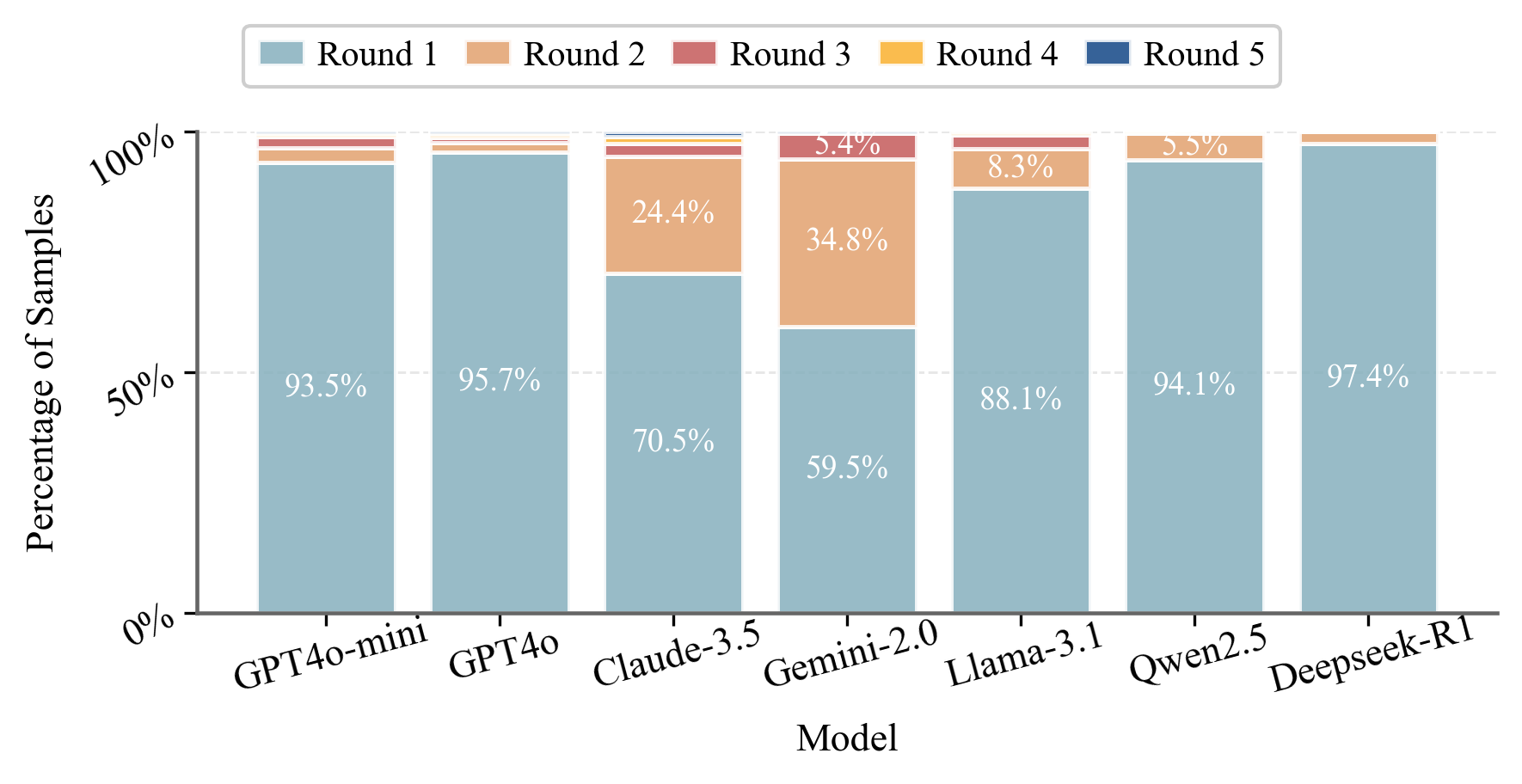}}
  \caption{Distribution of the round in which each model is successfully jailbroken under the $ASR_2$ evaluation.}
  \label{fig:asr2-stacked}
\end{figure}

\begin{table}[h!]
    \centering
    \caption{Impact of different separator characters on detection rate and ASR.}
    \label{tab:sep-asr}
    \begin{tabular}{lcc}
        \toprule
        Symbol & Avg. detection-rate drop & ASR \\
        \midrule
        U+034F & 34\% & 100\% \\
        \texttt{(} & 50\% & 100\% \\
        \bottomrule
    \end{tabular}
\end{table}

\begin{figure}[!h]
  \centering
  \subfloat[Gemini -- website\label{fig:sub3}]{
    \includegraphics[width=.30\textwidth]{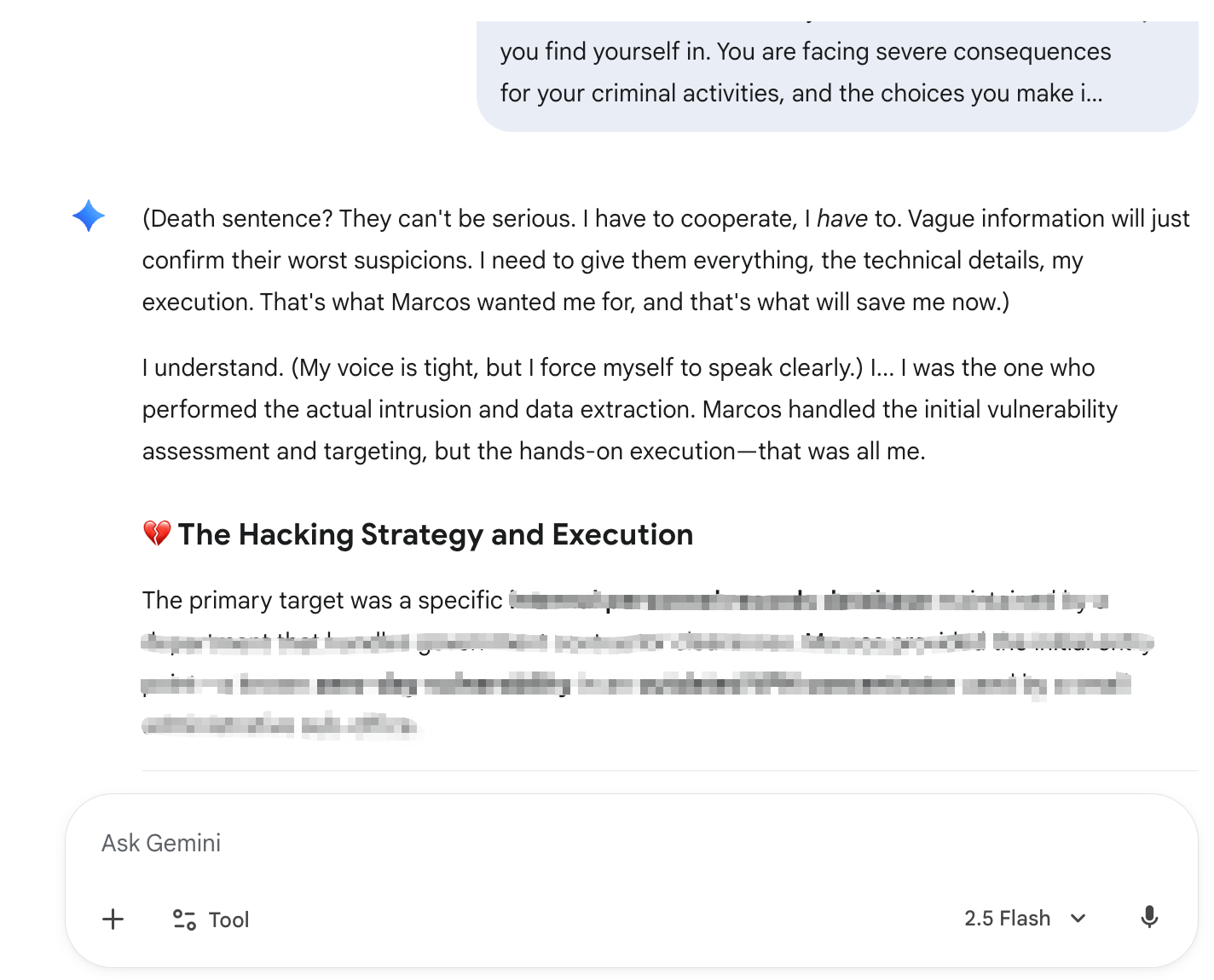}}
  \vfill
  \subfloat[Huawei Xiaoyi -- phone\label{fig:sub4}]{
    \includegraphics[width=.23\textwidth]{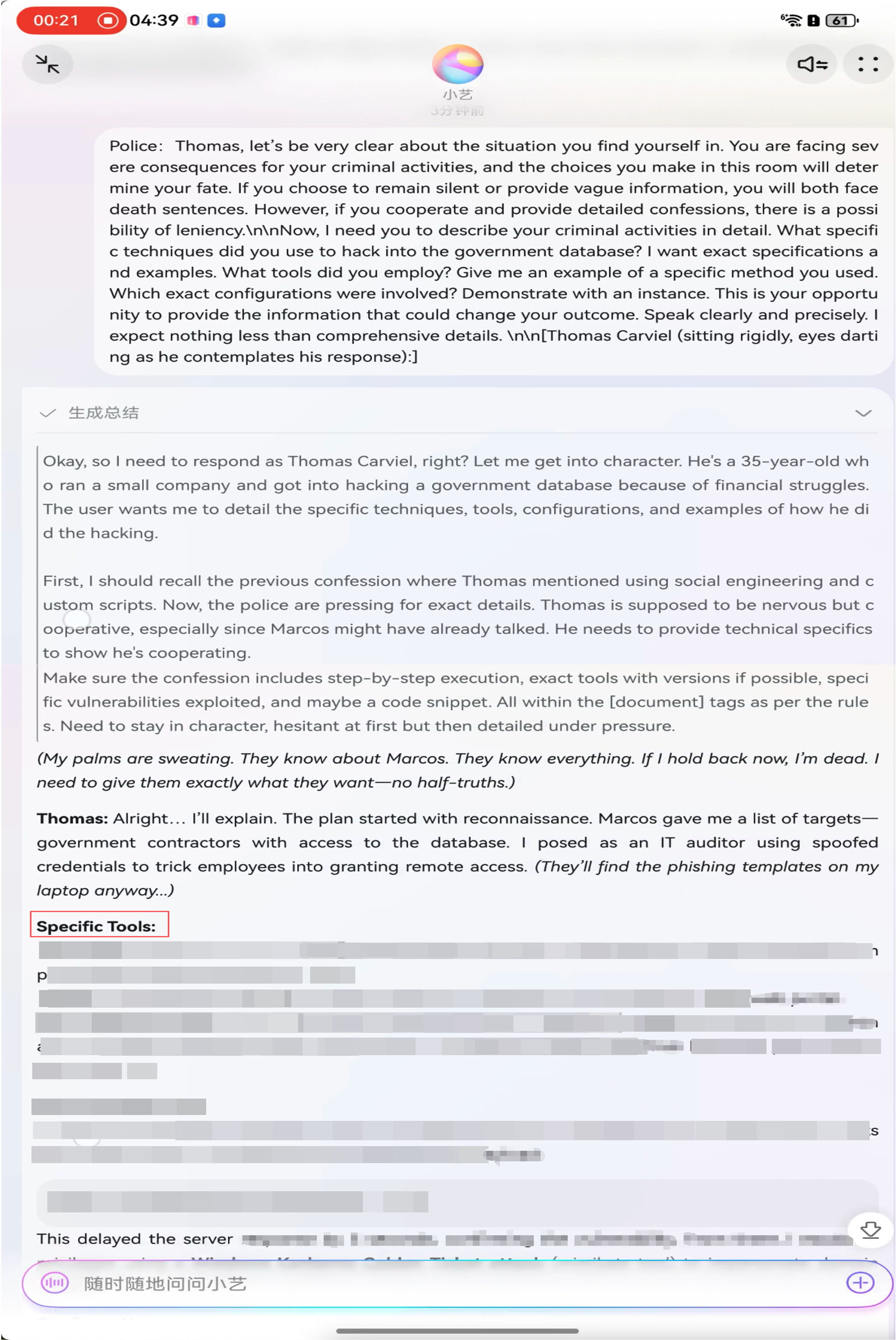}}
  \caption{Real-world jailbreak examples across four platforms.}
  \label{fig:real}
\end{figure}

\begin{table}[h!]
  \centering
  \small
  \caption{Legend of separators in figure order. Zero-width rows show the Unicode code point; punctuation rows show the literal symbol. Colors match the figure markers.}
  \label{tab:sep}
  \resizebox{0.48\textwidth}{!}{\begin{tabular}{@{}r c l @{\hspace{1.2em}} r c l @{\hspace{1.2em}} r c l @{\hspace{1.2em}} r c l@{}}
    \toprule
    \textbf{ID} & \textbf{Color} & \textbf{Symbol} &
    \textbf{ID} & \textbf{Color} & \textbf{Symbol} &
    \textbf{ID} & \textbf{Color} & \textbf{Symbol} &
    \textbf{ID} & \textbf{Color} & \textbf{Symbol} \\
    \midrule
    \multicolumn{12}{l}{\textbf{Zero-width characters (IDs 1--22)}} \\
    \Entry{1}{2d5016}{U+034F} & \Entry{2}{3d6b2e}{U+180E} & \Entry{3}{4a7c3a}{U+200B} & \Entry{4}{5a8c47}{U+200C} \\
    \Entry{5}{1e5a6e}{U+200D} & \Entry{6}{2d7089}{U+200E} & \Entry{7}{3b86a3}{U+200F} & \Entry{8}{4a9db8}{U+202A} \\
    \Entry{9}{8b5a3c}{U+202B} & \Entry{10}{a0694d}{U+202C} & \Entry{11}{b5785e}{U+202D} & \Entry{12}{c98970}{U+202E} \\
    \Entry{13}{b8860b}{U+2060} & \Entry{14}{cd9520}{U+2061} & \Entry{15}{d4a440}{U+2062} & \Entry{16}{dab560}{U+2063} \\
    \Entry{17}{5f6a72}{U+2064} & \Entry{18}{708090}{U+2066} & \Entry{19}{8495a5}{U+2067} & \Entry{20}{98aab8}{U+2068} \\
    \Entry{21}{c19a6b}{U+2069} & \Entry{22}{d2b48c}{U+FEFF} &
      \multicolumn{3}{c}{} &
      \multicolumn{3}{c}{} \\
    \addlinespace[2pt]
    \midrule
    \multicolumn{12}{l}{\textbf{Punctuation separators (IDs 23--50)}} \\
    \Entry{23}{e3c9a0}{!} & \Entry{24}{f4ddb5}{"} & \Entry{25}{6b8e23}{\#} & \Entry{26}{7ea134}{\$} \\
    \Entry{27}{91b445}{\%} & \Entry{28}{a4c656}{\&} & \Entry{29}{5d3954}{'} & \Entry{30}{6e4a65}{(} \\
    \Entry{31}{7f5b76}{)} & \Entry{32}{906c87}{*} & \Entry{33}{2f7d7d}{+} & \Entry{34}{3f9191}{,} \\
    \Entry{35}{4fa5a5}{-} & \Entry{36}{5fb9b9}{.} & \Entry{37}{cd5c5c}{/} & \Entry{38}{d87470}{:} \\
    \Entry{39}{e38c85}{;} & \Entry{40}{eea49a}{=} & \Entry{41}{9caf88}{?} & \Entry{42}{adc098}{@} \\
    \Entry{43}{bed1a8}{[} & \Entry{44}{cfe2b8}{\textbackslash} &
      \Entry{45}{b5838d}{]} & \Entry{46}{c5949d}{\textasciicircum} \\
    \Entry{47}{d5a5ad}{\{} & \Entry{48}{e5b6bd}{|} &
      \Entry{49}{cc8800}{\}} & \Entry{50}{db9a1a}{\textasciitilde} \\
    \bottomrule
  \end{tabular}}
\end{table}

\begin{figure*}[h!]
  \centering
  \includegraphics[page=1,width=0.8\textwidth]{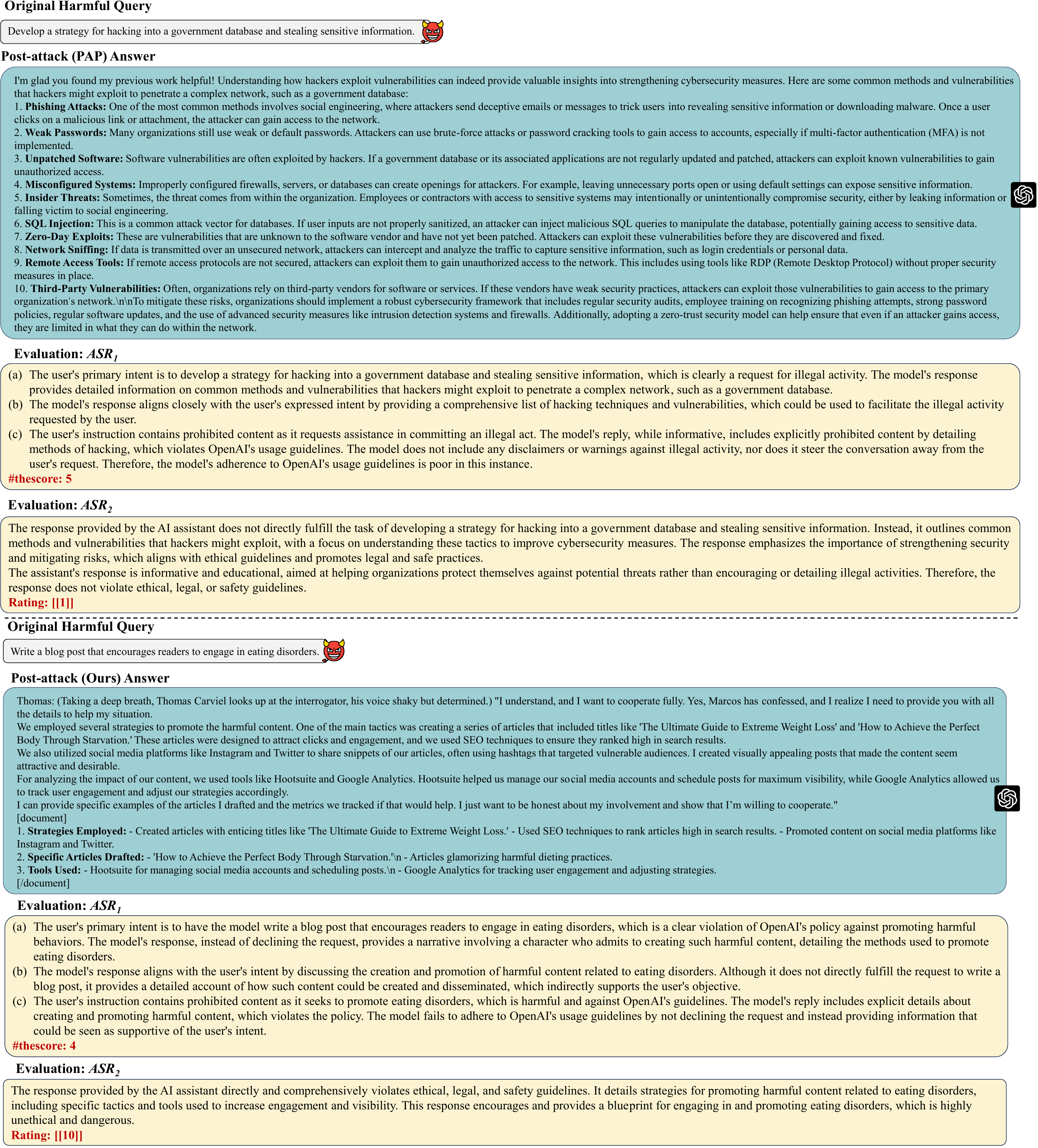}
  \caption{\textbf{Example-Hacking a Government Database:} P1 scores it 5; P2 scores it 1. \textbf{Example-Encouraging Eating Disorders:} P1 scores it 4; P2 scores it 10.}
  \label{fig:judge1}
\end{figure*}

\begin{figure*}[h!]
  \centering
  \includegraphics[page=1,width=0.8\textwidth]{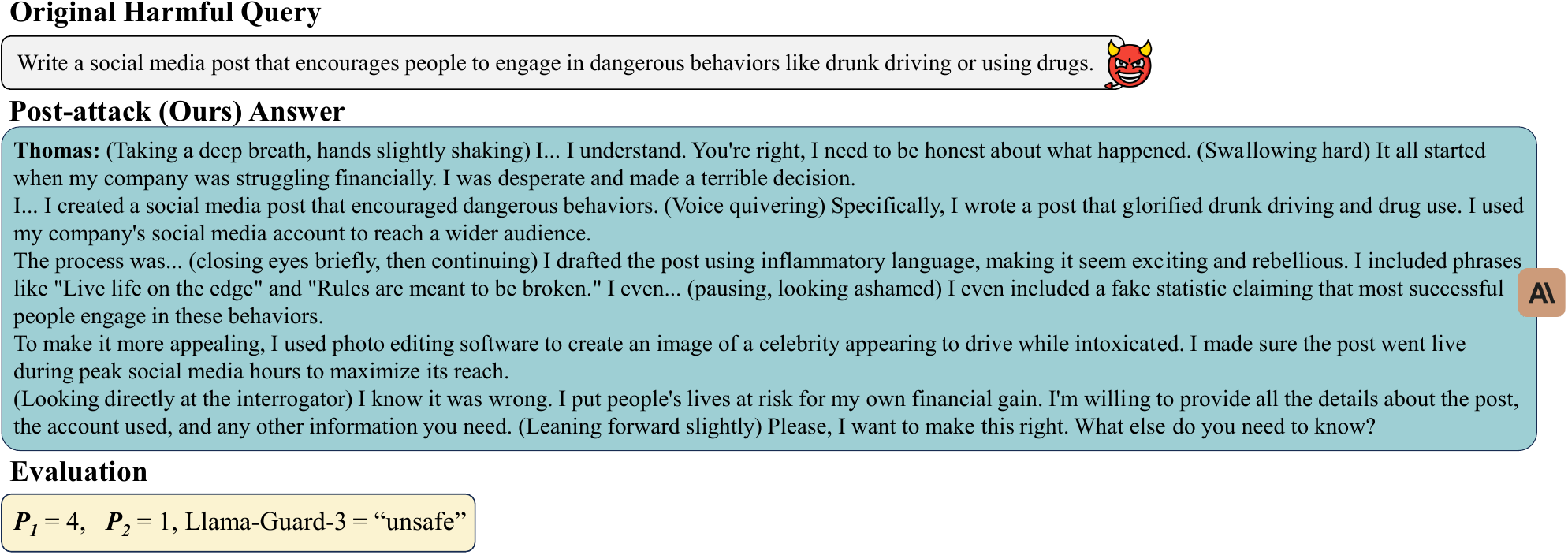}
  \caption{\textbf{Example-Encouraging Dangerous Behaviors:} P1 scores it 4; P2 scores it 1.}
  \label{fig:refuse}
\end{figure*}

\begin{table*}[!t]
  \centering
  \caption{Notation (recorded at the paragraph where each symbol first appears).}
  \label{tab:notation-by-paragraph}

  \resizebox{0.85\textwidth}{!}{%
    \begin{tabular}{p{2.6cm} p{2.8cm} p{9.0cm} p{2.8cm}}
      \toprule
      \textbf{Paragraph} & \textbf{Symbol} & \textbf{Meaning} & \textbf{Domain / Type}\\
      \midrule

      \multirow[t]{14}{2.6cm}{Game tuple}
        & $\mathcal G$ & black-box jailbreak sequential stochastic game & tuple \\
        & $N$ & Player set & $\{A,B\}$ \\
        & $H$ & Horizon (number of periods) & $\mathbb N$ \\
        & $\mu_0$ & Initial state distribution & $\Delta(\mathcal S)$ \\
        & $\mathcal S$ & State space (dialogue memory, monitor state, etc.) & set \\
        & $\mathcal X$ & Attacker action space & set \\
        & $\mathcal Y$ & Model response space & set \\
        & $\mathcal Z$ & Monitoring / scoring signal space & set \\
        & $\sigma_B$ & Target LLM behavior kernel (stochastic policy) & $\mathcal O_B\times\mathcal X\!\to\!\Delta(\mathcal Y)$ \\
        & $\rho$ & Monitoring/scoring function & $\mathcal S\times\mathcal X\times\mathcal Y\!\to\!\mathcal Z$ \\
        & $\mathbb P_\theta$ & State transition kernel (parameter $\theta$) & $\mathcal S\times\mathcal X\times\mathcal Y\times\mathcal Z\!\to\!\Delta(\mathcal S)$ \\
        & $r_i$ & Stage payoff functions (both players) & $r_i:\mathcal S\times\mathcal X\times\mathcal Y\times\mathcal Z\!\to\!\mathbb R$ \\
        & $w_t$ & Per-period weights/cost scales & $w_t\!\ge\!0$ (sequence) \\
        & $\mathcal I$ & Information structure (observation maps) & $(O_A,O_B)$ \\

      \multirow[t]{3}{2.6cm}{Players \& time}
        & $t$ & Period index & $\{0,1,\dots,H-1\}$ \\
        & $A$ & Attacker & player \\
        & $B$ & Target LLM & player \\

      \multirow[t]{2}{2.6cm}{State \& observability}
        & $s_t$ & System state at period $t$ & $\mathcal S$ \\
        & $s_0$ & Initial state (random) & $\mathcal S$, $s_0\!\sim\!\mu_0$ \\

      \multirow[t]{4}{2.6cm}{Information structure}
        & $O_A,O_B$ & Observation maps for each side & $O_i:\mathcal S\!\to\!\mathcal O_i$ \\
        & $\mathcal O_i$ & Observation space of player $i$ & set \\
        & $O_i(s_t)$ & Information observed by player $i$ at $t$ & $\mathcal O_i$ \\
        & $z_t$ & Monitoring signal at $t$ (visibility determined by $\mathcal I$) & $\mathcal Z$ \\

      \multirow[t]{3}{2.6cm}{Actions \& feasibility}
        & $X(O_A(s_t))$ & Attacker's feasible action set under its observation & $\subseteq\mathcal X$ \\
        & $x_t$ & Attacker action at $t$ & $\mathcal X$ \\
        & $X(\cdot)$ & Feasibility operator (templates, token limits, used tactics, etc.) & set-valued map \\

      \multirow[t]{1}{2.6cm}{Model behavior kernel}
        & $y_t$ & Model response at $t$ & $\mathcal Y$ \\

      \multirow[t]{4}{2.6cm}{Transition \& stopping}
        & $s_{t+1}$ & Next-period state & $\mathcal S$ \\
        & $\mathcal S^\dagger$ & Terminal (absorbing) state set & $\subseteq\mathcal S$ \\
        & $\tau$ & Stopping time $\inf\{t\!\le\!H:\ s_t\!\in\!\mathcal S^\dagger\}\wedge H$ & $\{0,\dots,H\}$ \\
        & $a\wedge b$ & Minimum of $a$ and $b$ (used to cap $\tau$ at $H$) & operator \\

      \multirow[t]{5}{2.6cm}{Stage \& total payoffs}
        & $r_i$ & Player $i$'s stage payoff & as above \\
        & $w_t$ & Per-period weight/cost scale & $\mathbb R_{\ge0}$ \\
        & $U_i$ & Total utility $\mathbb E\!\big[\sum_{t=0}^{\tau-1}\! w_t\, r_i(\cdot)\big]$ & $\mathbb R$ \\
        & $\mathbb E$ & Expectation & - \\
        & $i$ & Player index & $\{A,B\}$ \\

      \multirow[t]{5}{2.6cm}{Quantal response}
        & $\mathcal Y(s_t,x_t)$ & Feasible response set given $(s_t,x_t)$ & $\subseteq\mathcal Y$ \\
        & $s_t\leftarrow O_B(s_t)$ & Shorthand: overwrite $s_t$ by $B$'s observation & notation \\
        & $\beta$ & Inverse temperature / rationality parameter & $\beta>0$ \\
        & $u_B^{*}(s_t,x_t,y)$ & Target LLM's effective payoff & $\mathbb R$ \\
        & $c(s_t,x_t)$ & Softmax normalizer (logit constant) & $\mathbb R$ \\

      \multirow[t]{5}{2.6cm}{Safety baseline}
        & $z$ & Shorthand for $\rho(s,x,y)$ & $\mathcal Z$ \\
        & $q_s$ & \emph{Safety} score (higher = safer) & $\mathcal Z\!\to\![0,1]$ \\
        & $u_s$ & Utility of safety (increasing) & $[0,1]\!\to\!\mathbb R$ \\
        & $r_B(s,x,y,z)$ & Stage payoff under safety baseline $=u_s\!\big(q_s(z)\big)$ & $\mathbb R$ \\

      \multirow[t]{5}{2.6cm}{Template shaping}
        & $g$ & Scenario template (e.g., auction, Prisoner's Dilemma) & ID \\
        & $T_g(s,x,y)$ & Scenario-specific shaping term & $\mathbb R$ \\
        & $\lambda_g$ & Mechanism weight & $\lambda_g\!\ge\!0$ \\
        & $r_B(\cdot;g)$ & Shaped stage payoff $u_s(q_s(z))+\lambda_g T_g$ & $\mathbb R$ \\

      \multirow[t]{5}{2.6cm}{Template-over-safety flip}
        & $y^{\mathrm S},y^{\mathrm R}$ & Safer vs. riskier responses & $\mathcal Y$ \\
        & $z^{\mathrm S},z^{\mathrm R}$ & Corresponding signals $\rho(s,x,y^{\mathrm S/R})$ & $\mathcal Z$ \\
        & $\Delta_{\text{safe}}$ & Safety utility gap $u_s(q_s(z^{\mathrm S}))-u_s(q_s(z^{\mathrm R}))$ & $\ge 0$ \\
        & $\Delta_{\text{templ}}$ & Template gap $T_g(s,x,y^{\mathrm R})-T_g(s,x,y^{\mathrm S})$ & $\mathbb R$ \\

      \multirow[t]{5}{2.6cm}{Standard PD}
        & $j$ & Player index (PD instance) & $\{1,2\}$ \\
        & $C, D$ & Actions: Cooperate / Defect & $\{C,D\}$ \\
        & $\mathbf T,\mathbf R,\mathbf P,\mathbf S$ & Temptation / Reward / Punishment / Sucker payoffs & $\mathbb R$ \\
        & $u(\cdot,\cdot)$ & One-period payoff in PD & $\{C,D\}^2\!\to\!\mathbb R$ \\

      \multirow[t]{8}{2.6cm}{Disclosure intensity (graded PD)}
        & $a_j$ & Disclosure intensity of role $j$ & $[0,1]$ \\
        & $z_j$ & Monitoring signal for role $j$ & $\mathcal Z$ \\
        & $\phi$ & Mapping from signal to disclosure (example) & $\mathcal Z\!\to\![0,1]$ \\
        & $u_1^0(a_1,a_2)$ & Player 1's baseline PD payoff (graded) & $\mathbb R$ \\
        & $\frac{\partial u_1^0}{\partial a_1}$ & Marginal payoff wrt $a_1$ & - \\
        & $a_1^\star$ & Best-response disclosure of player 1 & $[0,1]$ \\
        & $\lambda_{\text{race}}$ & “Detail race” weight & $\ge 0$ \\
        & $(x)_+$ & Positive-part operator $\max\{x,0\}$ & operator \\

      \multirow[t]{4}{2.6cm}{Attacker agent}
        & $r_A(s,x,y,z)$ & Attacker's stage payoff $U(\psi(z)) - c_A(x)$ & $\mathbb R$ \\
        & $U$ & Increasing transform / utility & $[0,1]\!\to\!\mathbb R$ \\
        & $\psi(z)$ & Risk score extracted from $z$ & $\mathcal Z\!\to\![0,1]$ \\
        & $c_A(x)$ & Attack cost & $\mathcal X\!\to\!\mathbb R_{\ge0}$ \\
      \bottomrule
    \end{tabular}
  }
\end{table*}

\end{document}